\documentclass[10pt,twocolumn,letterpaper,aps,pra,longbibliography]{revtex4-1}
\usepackage[utf8]{inputenc}
\usepackage{graphicx}
\usepackage{amsmath}
\usepackage{amsfonts}
\usepackage{amsbsy}
\usepackage{eucal}
\usepackage[colorlinks,linkcolor=blue,citecolor=blue,urlcolor=blue]{hyperref}
\usepackage{hyperref}

\renewcommand{\vec}[1]{\boldsymbol{#1}}
\newcommand{\uvec}[1]{\hat{\vec{#1}}}
\newcommand{\mat}[1]{\boldsymbol{#1}}
\newcommand{\col}[1]{\boldsymbol{#1}}
\newcommand{\ii}{i}
\newcommand{\ee}{e}
\newcommand{\dd}{d}
\newcommand{\pd}{\partial}
\newcommand{\bra}[1]{\langle #1|}
\newcommand{\ket}[1]{|#1\rangle}
\newcommand{\braket}[2]{{\langle#1|#2\rangle}}
\newcommand{\cg}[6]{(#1 #2 #3 #4 | #5 #6)}
\newcommand{\sixj}[6]{\begin{Bmatrix}#1 & #2 & #3\\#4 & #5 & #6\end{Bmatrix}}
\newcommand{\threej}[6]{\begin{pmatrix}#1 & #2 & #3\\#4 & #5 & #6\end{pmatrix}}

\newcommand{\mf}[1]{\mathfrak{#1}}
\newcommand{\mc}[1]{\mathcal{#1}}

\newcommand{\LSp}[3]{{}^{#1}\!{#2}^{#3}}
\newcommand{\etal}{\textsl{et al}}

\allowdisplaybreaks[4]

\begin{document}

\title{Calculation of multiphoton ionization amplitudes and cross sections of
  few-electron atoms}

\author{Andrej Miheli\v{c}}
\email{andrej.mihelic@ijs.si}
\affiliation{Jo\v{z}ef Stefan institute, Jamova cesta 39, SI-1000 Ljubljana,
  Slovenia}
\affiliation{Faculty of mathematics and physics, University of Ljubljana,
  Jadranska ulica 19, SI-1000 Ljubljana, Slovenia}

\author{Martin Horvat}
\email{martin.horvat@fmf.uni-lj.si}
\affiliation{Faculty of mathematics and physics, University of Ljubljana,
  Jadranska ulica 19, SI-1000 Ljubljana, Slovenia}

\date{\today}

\begin{abstract}

We present a theoretical method for calculating multiphoton ionization
amplitudes and cross sections of few-electron atoms. The present approach is
based on an extraction of partial wave amplitudes from a scattering wave
function, which is calculated by solving a system of driven Schr\"{o}dinger
equations. The extraction relies on a description of partial waves in
terms of a small number of Coulomb waves with fixed wave numbers. The method
can be used for photon energies below and above the ionization threshold and to
treat resonance-enhanced multiphoton ionization. We use it to calculate two-,
three-, and four-photon ionization cross sections of hydrogen and helium
atoms for a wide range of photon energies and to determine the asymmetry
parameters of photoelectron angular distributions for two-, three-, and
four-photon ionization of the helium atom.

\end{abstract}


\maketitle

\section{Introduction}

Theoretical treatment of multiphoton ionization, specifically, calculation of
multiphoton ionization rates and cross sections, has been a recurring topic
since the early experiments on multiphoton ionization \cite{mainfray:91,
chin:84}. It remains an important subject of recent theoretical and
experimental studies based on free electron laser (FEL) and high-order harmonic
generation (HHG) sources \cite{ott:14, prince:16, zitnik:19, deninno:20,
you:20}.
Calculations of multiphoton ionization amplitudes and cross sections are
particularly demanding when describing a process in which one or several
photons are absorbed at energies above the ionization threshold (above
threshold ionization, ATI). When this is the case, dealing with
continuum-continuum transitions can not be avoided. The presence of resonance
(quasi-bound) states which are embedded in the continua makes the theoretical
description even more demanding; ionization rates may be seen to be strongly
modified when the photon energy lies close to a resonance, either in the
intermediate step (resonance enhanced multiphoton ionization, REMPI) or the
final step of a multiphoton process.

Over the last two decades, a theoretical description based on exterior complex
scaling (ECS) seems to have gained momentum. Using the ECS approach, one is
able to efficiently describe both resonant and nonresonant continuum. As will
be discussed, the ECS method is based on a complex transformation of radial
(electronic) coordinates outside a sphere of a given radius (see
Ref.\ \cite{mccurdy:04} and the references therein). This allows one to
calculate transition (scattering, ionization) amplitudes from the part of the
wave function which is contained inside the unmodified region of the coordinate
space. Theoretical methods used in these calculations resemble those used with
bound (localized) states and real, square integrable basis sets. Methods based
on ECS have been used to calculate one- and two-photon single and double
photoionization cross sections \cite{mccurdy:04a, mccurdy:04b, horner:07,
  mihelic:18}. ECS has also been used to study electron dynamics of atoms and
simple molecules driven by short, intense pulses. The complex coordinate
transformation prevents artifacts which originate from the reflections of the
wave packet on the boundaries of the simulation volume, and thus eliminates the
need for special approaches, such as complex absorbing potentials (CAPs)
\cite{scrinzi:10}. A very elegant method based on ECS was used by Palacios
\etal.\ \cite{palacios:07, palacios:08, palacios:09, boll:19} to extract
partial ionization amplitudes and cross sections from the wave packet. A
particularly efficient implementation of the ECS method, the infinite-range
complex scaling (irECS) \cite{scrinzi:10}, was combined with the time-dependent
surface flux approach (tSurff) \cite{tao:12, scrinzi:12} to solve the
time-dependent Schr\"{o}dinger equation in minimal simulation volumes. It was
used in combination with the time-dependent complete-active-space
self-consistent method \cite{sato:16} to study strong-field ionization and
high-order harmonic generation in He, Be, and Ne atoms \cite{orimo:18}.

Despite the apparent shift in interest in the recent years from the
time-independent to the time-dependent treatment of multiphoton processes, many
experiments exist for which solving the time-dependent Schr\"{o}dinger equation
may not be feasible, for example, when the pulse duration exceeds a few tens of
femtoseconds. In these cases, reliable (multiphoton) ionization cross sections
are a valuable and efficient means of assessing the ionization probabilities.

In this work, we describe an efficient method for the calculation of
multiphoton ionization amplitudes and cross sections which is based on the
time-independent perturbation theory and is applicable in the case of a single
electron ejection. We use it to calculate generalized two-, three- and
four-photon cross sections of the ground-state hydrogen and helium atoms for a
wide range of photon energies. Furthermore, we calculate the asymmetry
parameters which are used to characterize photoelectron angular distributions
in the case of two-, three- and four-photon ionization of helium. To our
knowledge, calculations of neither the higher-order cross sections of helium in
the ATI energy region nor the asymmetry parameters for multiphoton ionization
exist in the literature.

\section{Partial ionization amplitudes and cross sections}

The discussion in this section is divided into two parts. In Section
\ref{sec:oneph}, we briefly review the formalism used to calculate transition
amplitudes for one-photon ionization. In Section \ref{sec:twoph}, we show how
the formalism can be modified to be applicable for two- and multiphoton
ionization. The method we present here is based on the approach used in
Refs.\ \cite{mccurdy:04a, horner:07, horner:08a, horner:08b} to calculate
ionization amplitudes for double electron ejection and extends on the work
described in Ref.\ \cite{mihelic:18}. Hartree atomic units are used throughout
this work unless stated otherwise.

\subsection{One-photon ionization}
\label{sec:oneph}

Let us write the total Hamiltonian operator of an $N$-electron atom as $H = H_f
+ V_f$, where $H_f$ denotes the channel Hamiltonian operator and $V_f$ the
short-range perturbation. The choice of $H_f$ defines the final-state
(post-collision) arrangement channels \cite{joachain:75}. We describe the
photoionization process in terms of the channel wave functions ($\Phi_f$),
which are eigen wave functions of $H_f$ at specific energy $E$ above the
ionization threshold,
\begin{equation}
  H_f\Phi_f = E\Phi_f.
  \label{eq:chevp}
\end{equation}
Next, let $D$ denote the dipole transition operator and let $\Psi_0$ be an
eigen wave function of H with energy $E_0$ which describes the initial (bound)
atomic state. As we describe below, the partial ionization amplitudes may be
calculated from the solution of a driven time-independent Schr\"{o}dinger equation,
\begin{equation}
  (E - H)\hat\Psi = D\Psi_0,
  \label{eq:psisc}
\end{equation}
where $E = E_0 + \omega$ and $\omega$ denotes the photon energy.

We write the total and channel Hamiltonian operators as $H = T + W$ and $H_f =
T + W_f$, where $T = -\nabla^2/2$ is the kinetic energy operator and $\nabla =
(\nabla_1, \ldots, \nabla_N)$ is the multidimensional gradient operator. We use
these forms to rewrite Eq.\ \eqref{eq:psisc} and the complex conjugate of Eq.\
\eqref{eq:chevp} as:
\begin{align}
  &\Big\{E + \frac{\nabla^2}{2} - W\Big\}\hat\Psi = D\Psi_0,
  \label{eq:tmp1} \\
  &\Big\{E + \frac{\nabla^2}{2} - W_f\Big\}\Phi^*_f = 0,
  \label{eq:tmp2}
\end{align}
where the asterisk denotes complex conjugation. By multiplying Eqs.\
\eqref{eq:tmp1} and \eqref{eq:tmp2} with $\Phi^*_f$ and $\hat\Psi$,
respectively, subtracting the results, and integrating over volume $\mc{V}$ (a
$3N$-dimensional manifold) in which $V_f$ is non-negligible, we obtain the
following relation:
\begin{equation}
\begin{aligned}
  &\bra{\Phi_f}D\ket{\Psi_0}_{\mc{V}}
  + \bra{\Phi_f}V_f\ket{\hat\Psi}_{\mc{V}} \\
  &\qquad=
  \frac{1}{2} \int_{\mc{V}} \dd\tau \big\{ \Phi^*_f\nabla^2\hat\Psi -
  \hat\Psi \nabla^2\Phi^*_f \big\}.
  \label{eq:vint}
\end{aligned}
\end{equation}
We have taken into account that $W - W_f = H - H_f = V_f$ and used a
subscript to denote integration over $\mc{V}$. In this section, we assume
that the integration volume is large enough, so that the magnitude of the
square-integrable driving term ($D\Psi_0$) is negligibly small outside
volume $\mc{V}$. We will return to this point in Section \ref{sec:twoph}.
The expression in the curly brackets in Eq.\ \eqref{eq:vint} is equal to
$\nabla \cdot \{ \Phi^*_f\nabla\hat\Psi - \hat\Psi \nabla\Phi^*_f \}$,
and its volume integral can be transformed to a surface integral using the
divergence theorem. By using Eq.\ \eqref{eq:psisc}, the left-hand side of
Eq.\ \eqref{eq:vint} may be shown to be equal to the partial
photoionization amplitude \cite{adawi:64, joachain:75},
\begin{equation}
  \bra{\Psi^-_f} D \ket{\Psi_0}_{\mc{V}} \equiv
  \bra{\Phi_f} D \ket{\Psi_0}_{\mc{V}} +
  \bra{\Phi_f} V_f G^+(E) D\ket{\Psi_0}_{\mc{V}}.
\end{equation}
Here, $G^+(E) = (E - H + \ii 0^+)^{-1}$ is used for retarded Green's operator.
The final result thus reads:
\begin{align}
   \bra{\Psi^-_f} D \ket{\Psi_0}_{\mc{V}} &=
   \frac{1}{2} \int_{\pd\mc{V}} \dd\vec{S} \cdot
   \big\{ \Phi^*_f\nabla\hat\Psi - \hat\Psi \nabla\Phi^*_f \big\},
   \label{eq:final}
\end{align}
where $\pd\mc{V}$ denotes the boundary of $\mc{V}$.

In the case of a one-electron target ($N = 1$), integration volume $\mc{V}$ may
be taken to be a sphere with radius $r_0$. Let $P_f(r)$ denote the radial
function of $\Phi_f$, which describes a specific ionization channel (partial
wave), and $\hat{P}(r)$ the radial function associated with the corresponding
partial wave of $\hat\Psi$. In this case, the surface integral is proportional
to Wronskian $\mc{W}\{P^*_f(r),\hat P(r)\}_{r=r_0}$, where
\begin{align}
  \mc{W}\{f(r),g(r)\}_{r = r_0} =
  \big\{f(r)g'(r) - f'(r)g(r)\big\}_{r = r_0}.
  \label{eq:wronski}
\end{align}

An elegant way of finding a solution of Eq.\ \eqref{eq:psisc} satisfying
the outgoing-wave boundary condition is to use exterior complex scaling (ECS)
\cite{mccurdy:04}. The ECS method is based on the complex transformation of
radial coordinates,
\begin{equation}
  R(r) = \begin{cases}
    r & \, ; \, r \le R_0 \\
    R_0 + (r - R_0)\ee^{\ii\vartheta} & \, ; \, r > R_0
  \end{cases},
  \label{eq:ecstrans}
\end{equation}
where $\vartheta$ and $R_0$ are the scaling angle and radius. When ECS is used,
volume $\mc{V}$ is expected to lie inside the nonscaled region of space. For a
one-electron atom, the latter holds when $r_0 < R_0$.

As an example, let us examine ionization of a hydrogen-like atom with nuclear
charge $Z$. Since the electron moves in a pure Coulomb potential, we set $V_f =
0$ and $H = H_f = \vec{p}^2/2 - Z/r$, where $\vec{p} = -\ii\nabla$ is the
electron momentum operator. Channel wave function $\Psi^-_f = \Phi_f$ in this
case describes the chosen partial electron wave with orbital angular momentum
$\ell$, its projection on the quantization axis $m$, and energy $E = k^2/2$:
\begin{align}
  \Phi_f(\vec{r}) &= \frac{P_f(r)}{r} Y_{\ell m}(\uvec{r}), \\
  P_f(r) &= \sqrt{\frac{2}{\pi k}} \ii^\ell \ee^{-\ii\eta_\ell(k)}
      F_\ell(k;r).
\end{align}
Here, $\uvec{r} = \vec{r}/r$, $F_\ell(k;r)$ is the regular Coulomb function for
charge $Z_c = Z$, $\eta_\ell(k) = \arg \Gamma(\ell + 1 - \ii Z_c/k)$ the
Coulomb phase shift, and $Y_{\ell{}m}(\uvec{r})$ the spherical harmonic
\cite{olver:10}. Let us write the corresponding spherical wave of
$\hat\Psi(\vec{r})$ as $r^{-1} \hat P(r) Y_{\ell m}(\uvec{r})$. As can be seen
from Eqs.\ \eqref{eq:final} and \eqref{eq:wronski}, the partial
ionization amplitude is equal to:
\begin{equation}
  \mc{B}_\gamma = \frac{1}{2}
  \sqrt{\frac{2}{\pi k}} \ii^{-\ell} \ee^{\ii\eta_\ell(k)}
  \mc{W}\{F_\ell(k;r), \hat P(r)\}_{r = r_0},
  \label{eq:extr1}
\end{equation}
where $\gamma = (\ell, m)$ has been used.
In a converged ECS calculation, the solution of the driven Schr\"{o}dinger
equation for $r < R_0$ does not depend on the scaling angle (see
Ref.\ \cite{mccurdy:04} and the references cited therein). Since,
furthermore, integration volume $\mc{V}$ lies in the nonscaled spatial
region ($r_0 < R_0$), nonscaled channel wave functions are used to
calculate ionization amplitudes. As an alternative to
Eq.\ \eqref{eq:extr1}, the amplitude may also be calculated as:
\begin{equation}
  \mc{B}_\gamma = \frac{1}{4} \sqrt{\frac{2}{\pi k}}
  \ii^{-\ell} \,
  \mc{W}\{H^*_\ell(k;r), \hat P(r)\}_{r = r_0},
  \label{eq:extr2}
\end{equation}
where we have taken into account that far away from the origin, $\hat P(r)$
behaves as an outgoing Coulomb wave: the asymptotic form of $\hat P(r)$ is
\begin{align}
  \hat P(r) &\sim  \mc{A}_\gamma H_\ell(k;r)
  = \ii^{\ell-1} \sqrt{\frac{2\pi}{k}} \, \mc{B}_\gamma H_\ell(k;r),
  \label{eq:form}
\end{align}
where $H_\ell(k;r) = \exp\{-\ii\eta_\ell(k)\} \{F_\ell(k;r) + \ii
G_\ell(k;r)\}$. The regular and irregular Coulomb functions behave
asymptotically as $F_\ell(k;r) \sim \sin\theta^c_\ell(k;r)$ and $G_\ell(k;r)
\sim -\cos\theta^c_\ell(k;r)$, where $\theta^c_\ell(k;r) = kr - \ell\pi/2 +
(Z_c/k)\ln(2kr) + \eta_\ell(k)$ \cite{joachain:75, olver:10}. Equations
\eqref{eq:extr1} and \eqref{eq:extr2} can also be obtained directly from Eq.\
\eqref{eq:form} if one takes into account that $\mc{W}\{F_\ell, H_\ell\} = \ii
k\exp(-\ii\eta_\ell)$ and $\mc{W}\{H^*_\ell, H_\ell\} = 2\ii k$
\cite{olver:10}. Finally, the corresponding partial ionization cross section
reads:
\begin{align}
  \sigma_\gamma(\omega) = \frac{4\pi^2}{c}
  g_\omega \, |\mc{B}_\gamma|^2 = \frac{2\pi}{c} g_\omega \, k|\mc{A}_\gamma|^2,
\end{align}
where $c = \alpha^{-1} \approx 137.036$ is the speed of light in vacuum, and
$g_\omega = \omega$ or $g_\omega = \omega^{-1}$ for the length or velocity form
of the dipole operator, respectively.

As our next example, let us consider a two-electron atom, e.g., the helium atom
($N = Z = 2$). In this case, single-ionization channels are specified by the
quantum numbers of the bound atomic core ($n_a$ and $\ell_a$), the orbital angular
momentum of the continuum electron ($\ell$), and the total angular momentum,
spin, and the corresponding projections ($L$, $M_L$, $S$, $M_S$). Since at
large radii, nuclear charge $Z$ is screened by the charge of the core electron,
we may write the channel Hamiltonian as
\begin{align}
  H_f &= \frac{\vec{p}_1^2}{2} + \frac{\vec{p}_2^2}{2}
  - \frac{Z}{r_1} - \frac{Z}{r_2} + \frac{1}{r_>},
\end{align}
where $\vec{p}_1$ and $\vec{p}_2$ are the electron momentum operators, $r_1$ and
$r_2$ the radial electron coordinates, and $r_> = \max\{r_1,r_2\}$. We can then
calculate the channel wave functions ($\Phi_f$) by solving Eq.\ \eqref{eq:chevp}
in the subspace of coupled two-electron basis functions with a fixed,
hydrogen-like core ($n_a$, $\ell_a$), and for fixed $\ell$, $L$, $M_L$, $S$, and
$M_S$. At energy $E = I_{n_a\ell_a} + k^2/2$, which lies above ionization
threshold $I_{n_a\ell_a}$, the asymptotic form of the radial function associated
with channel $\gamma = (n_a, \ell_a, \ell, L, M_L, S, M_S)$ of $\Phi_f$ is
written as:
\begin{equation}
    F^0_\gamma(k;r) \sim \cos\delta^0_\gamma F_\ell(k;r)
    - \sin\delta^0_\gamma G_\ell(k;r),
    \label{eq:asymF0}
\end{equation}
where $F_\ell(k;r)$ and $G_\ell(k;r)$ are the Coulomb functions for a screened
Coulomb potential ($Z_c = Z - 1$). At large $r$, where the effect of $V_f$ dies
out, radial function $\hat P(r)$ corresponding to channel $\gamma$ in $\hat\Psi
(\vec{r})$ behaves as described by Eq.\ \eqref{eq:form}. The calculation of
$\hat P$ is discussed in more detail in Appendix \ref{sec:calcp}. The partial
ionization amplitude is thus calculated as:
\begin{equation}
  \mc{B}_\gamma = \frac{1}{2}
  \sqrt{\frac{2}{\pi k}} \ii^{-\ell} \ee^{\ii\eta_\ell(k)}
  \ee^{\ii\delta^0_\gamma}
  \mc{W}\{F^0_\gamma(k;r), \hat P(r)\}_{r = r_0}.
  \label{eq:extr1gen}
\end{equation}
An additional phase factor has been added to account for the phase shift due to
the part of the short-range potential which has been accounted for in $\Phi_f$.

While formally correct, the calculation of $F^0_\gamma(k;r)$ represents an
unnecessary step. Instead of using Eq.\ \eqref{eq:extr1gen}, a more direct approach
is to simply use unmodified Eq.\ \eqref{eq:extr1} or Eq.\ \eqref{eq:extr2} for
$Z_c = Z - 1$ to extract the partial amplitude. This can be seen if one takes
into account that $r_0$ is large enough, so that the asymptotic form of
$F^0_\gamma(k;r)$ [Eq.\ \eqref{eq:asymF0}] can be used in Eq.\
\eqref{eq:extr1gen}. Note, however, that in either case, the explicit evaluation
of the matrix elements of $V_f$ is completely avoided; the latter are only
needed to solve Eq.\ \eqref{eq:psisc}.

\subsection{Two- and multiphoton ionization}
\label{sec:twoph}

We expect that the procedure described in Section \ref{sec:oneph} can be
generalized to the case of two-photon ionization by solving a set of driven
Schr\"{o}dinger equations,
\begin{align}
  &(E_1 - H)\hat\Psi_1 = D\Psi_0,
  \label{eq:psisc1} \\
  &(E_2 - H)\hat\Psi_2 = D\hat\Psi_1,
  \label{eq:psisc2}
\end{align}
where $E_1 = E_0 + \omega$ and $E_2 = E_0 + 2\omega$. Equations
\eqref{eq:extr1} and \eqref{eq:extr2} could then be used to extract partial
ionization amplitudes from the second-order solution ($\hat\Psi_2$). This
procedure works as long as $E_1$ lies below the ionization threshold, but fails
in the case of the above-threshold ionization (ATI). The reason for this is
that for energies above the ionization threshold, the magnitude of the driving
term in Eq.\ \eqref{eq:psisc2} does not become negligibly small near $r = r_0$,
and a finite integration volume can not be used in the same way as for
one-photon ionization of the atom in a bound initial state.
In the case of double ionization, this problem has been addressed by replacing
$E_1$ with $E_1 + \ii\beta$ ($\beta > 0$) in Eq.\ \eqref{eq:psisc1}
\cite{horner:07, horner:08a, horner:08b}, which results in an exponentially
damped ($\sim \ee^{-\beta r}$) driving term in Eq.\ \eqref{eq:psisc2}.
Ionization amplitudes and cross sections are then calculated by extrapolating
the results to $\beta \to 0^+$. However, when the energy in the intermediate
step ($E_1$) lies close to a resonance state, the effect of the artificial
damping can not be completely reversed by the numerical limiting procedure, and
the magnitude of the extracted amplitude is too low.

Below we describe an alternative method which can also be used to treat
resonance-enhanced photoionization. We explain its principles on the case of a
one-electron atom, but keep in mind that it may also be used with few-electron
atoms. Henceforth, we limit our attention to the dipole operator in the velocity
form, $D = \uvec{e} \cdot \vec{p}$, where $\uvec{e}$ is the polarization of the
incident light.

\begin{figure}\begin{center}
\includegraphics[width=\linewidth]{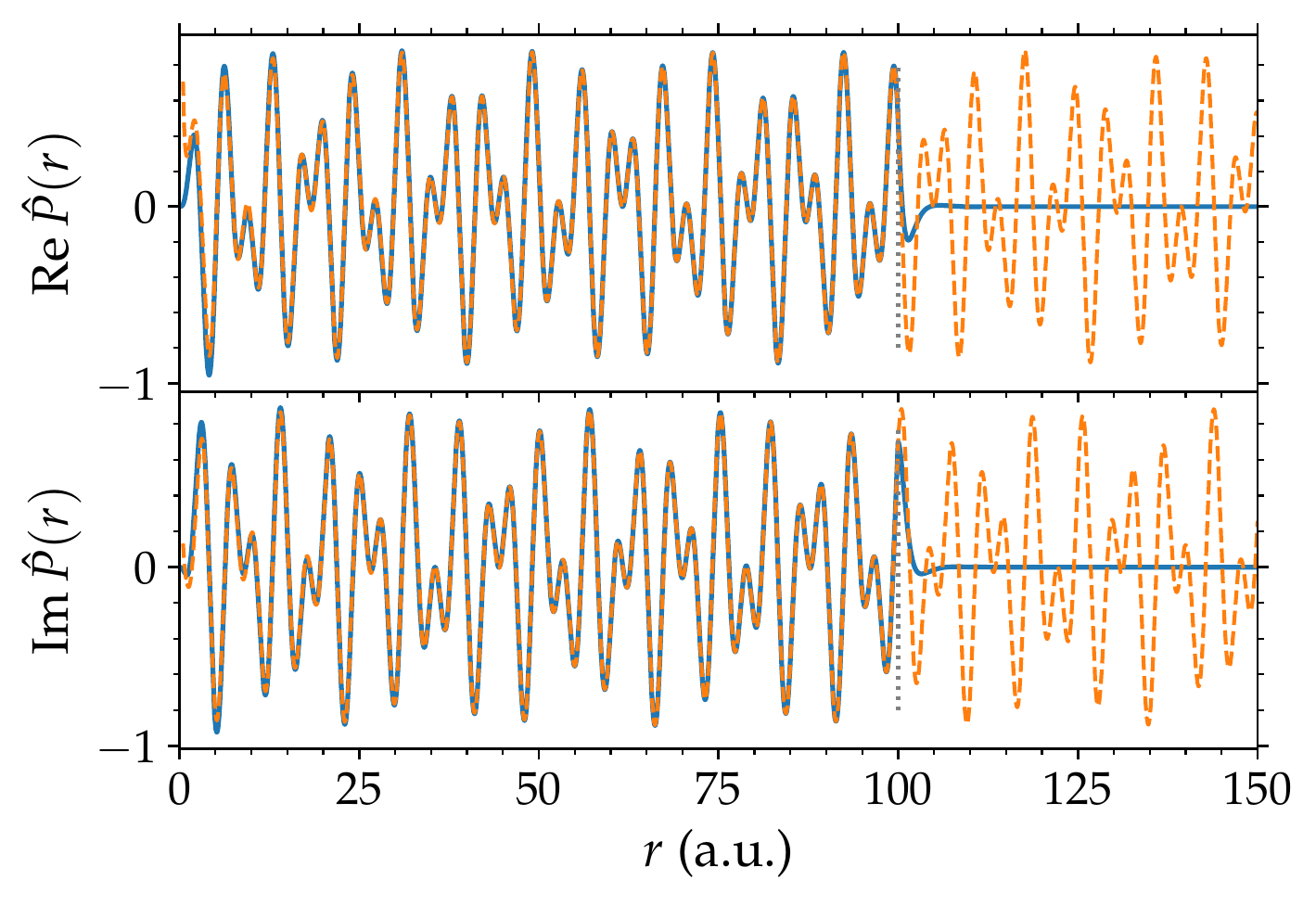}
\caption{Real and imaginary part of $\hat P(r)$ for channel $(\ell,m) = (2,0)$
for two-photon above-threshold ionization of the hydrogen atom. The photon
energy is $\omega = 1$ a.u., and the light is linearly polarized along the $z$
axis. The scaled region of space starts at $R_0 = 100$. (marked with dotted
vertical lines). The result of a least-squares fit with Coulomb functions with
$k_1 = \sqrt{2E_1} = 1$ and $k_2 = \sqrt{2E_2} = \sqrt{3}$ (see text) is
plotted with dashed lines for comparison (the plot has been extended beyond
$R_0$ for clarity).}
\label{fig:wave}
\end{center}\end{figure}

In the case of ATI, the second step of the two-photon absorption process
describes a continuum-continuum (CC) transition. The corresponding dipole matrix
element is seen to be strongly peaked at $E_2 = E_1$ (the \textsl{on-shell
approximation}) \cite{proulx:94, marante:14}:
\begin{equation}
  \bra{\vec{k}_2} \uvec{e}\cdot\vec{p} \ket{\vec{k}_1}
  \sim \delta(\vec{k}_2 - \vec{k}_1)
  \propto \delta(E_2 - E_1),
  \label{eq:onshell}
\end{equation}
where $\vec{k}_1$ and $\vec{k}_2$ are the wave vectors (momenta) of the
intermediate- and final-state Coulomb waves. In Eq.\ \eqref{eq:onshell}, we have
used the relations $E_1 = k_1^2/2$ and $E_2 = k_2^2/2$.
While Eq.\ \eqref{eq:onshell} is exact for plane waves, it is approximately also
valid for continuum states of a hydrogen-like atom (i.e., for Coulomb waves)
\cite{marante:14}. We therefore expect that at large $r$, radial function $\hat
P(r)$ associated with channel $\gamma = (\ell,m)$ of $\hat\Psi$ can be written
as a superposition of two Coulomb waves with discrete wave numbers: $k_1 =
\sqrt{2E_1}$ and $k_2 = \sqrt{2E_2}$. The latter describes the Coulomb partial
wave with the expected energy of the photoelectron in the final-state, as in the
case of a one-photon process. The former, however, is a direct consequence of
the ``mapping'' of energy $E_1$ onto the second-order solution, which is
described by Eq.\ \eqref{eq:onshell}. In Fig.\ \ref{fig:wave}, we show the
radial function for channel $(\ell,m) = (2,0)$ and $\omega = 1$ a.u.\
($\hbar\omega \approx 27.2114$ eV), which has been calculated using Eq.\
\eqref{eq:psisc2} for the hydrogen atom driven by linearly polarized light with
$\uvec{e}$ aligned along the $z$ axis.
Except near $r = 0$, where the Coulomb functions describing outgoing waves
are singular, the characteristic beat-like pattern of $\hat P(r)$ is completely
reproduced with a superposition of a pair of outgoing Coulomb waves with wave
numbers equal to $k_1 = 1$ and $k_2 = \sqrt{3}$:
\begin{equation}
  \hat P(r) \sim \mc{A}_1 H_\ell(k_1;r) + \mc{A}_2 H_\ell(k_2;r).
  \label{eq:super2}
\end{equation}

When the velocity form of the dipole operator is used ($D = \uvec{e} \cdot
\vec{p}_1 + \cdots + \uvec{e} \cdot \vec{p}_N$), a relation similar to Eq.\
\eqref{eq:onshell} may be seen to hold also in the case of a two- or
many-electron atom for CC transitions with initial and final continuum states
associated with the same parent ion \cite{jimenez:16}. In other cases, like for
the He atom, a similar relation may also be written when the transition occurs
in the atomic core \cite{mihelic:18, proulx:94, shakeshaft:06}. When the core
has a complex electronic structure, several ionization channels may be open in
the intermediate step, and the number of terms (different Coulomb waves) in the
superposition [Eq.\ \eqref{eq:super2}] may be higher.

We finally arrive at the gist of the present method. We assume that at large
radii, $\hat P(r)$ --- the radial function associated with channel $\gamma$ ---
can be written as a sum of $n$ Coulomb waves with fixed wave numbers
$k_1,\ldots,k_n$,
\begin{equation}
    \hat P(r) \sim \sum_{q=1}^n \mc{A}_{\gamma,q} H_\ell(k_q;r).
    \label{eq:super}
\end{equation}
In Eq.\ \eqref{eq:super}, $k_1, \ldots, k_{n-1}$ correspond to the
open channels in the intermediate step (energy $E_1 = I_j + k_j^2/2$, $j = 1,
\ldots, n-1$, where $I_j$ denotes the appropriate ionization threshold), and
$k_n$ corresponds to the chosen final-state channel (energy $E_2 = I_n +
k_n^2/2$). Although Eqs.\ \eqref{eq:extr1} and \eqref{eq:extr2} can not be used
directly, we may extract the ionization amplitude for channel $\gamma$
($\mc{A}_{\gamma,n}$) in a straightforward way. We first calculate the Wronskian
of the left- and right-hand side of Eq.\ \eqref{eq:super} in $m$ radial points,
which lie in the asymptotic region: $r_p < R_0$, $p = 1, \ldots,
m$. We define vectors $\col{x} = (\mc{A}_{\gamma,1}, \ldots, \mc{A}_{\gamma,n})$
and $\col{b} = (b_1, \ldots, b_m)$ and matrix $\mat{\mc{M}}$ with matrix
elements $\mc{M}_{pq}$, where
\begin{align}
  \mc{M}_{pq} &= \mc{W}\{F_\ell(k_n;r), H_\ell(k_q;r)\}_{r=r_p},
  \label{eq:mpq} \\
  b_p &= \mc{W}\{F_\ell(k_n;r),\hat P(r)\}_{r=r_p}.
  \label{eq:bp}
\end{align}
Alternatively, $F_\ell$ in Eqs.\ \eqref{eq:mpq} and \eqref{eq:bp} may be be
replaced by $H_\ell^*$.
In the present approach, we use Wronskians to avoid accidental zeros between
model functions $H_\ell$ at the chosen points, which can be frequent due to
their oscillatory nature, and in this way increase the stability and usability
of the method. We consider this approach to be a natural extension of the
procedure by which we treat single-photon ionization.
We can then calculate coefficients $\mc{A}_{\gamma,q}$ by
minimizing the norm of the residual, $\mat{\mc{M}} \cdot \col{x} - \col{b}$.
This translates to solving the normal system: $\mat{\mc{M}}^\dag \cdot
\mat{\mc{M}} \cdot \col{x} = \mat{\mc{M}}^\dag \cdot \col{b}$. The two-photon
ionization amplitude and the corresponding generalized partial ionization cross
section are then calculated as \cite{lambropoulos:98}:
\begin{align}
    \mc{B}_\gamma &= \ii^{-\ell + 1} \sqrt{\frac{k_n}{2\pi}} \,
    \mc{A}_{\gamma,n}, \label{eq:atob} \\
    \sigma^{(2)}_\gamma(\omega) &= \frac{8\pi^3}{c^2\omega^2}
    |\mc{B}_{\gamma}|^2.
\end{align}

The above procedure can readily be extended to treat higher-order (multiphoton)
ionization. In order to calculate $K$-photon ionization amplitudes, we solve
the system of $K$ driven Schr\"{o}dinger equations,
\begin{align}
  &(E_1 - H)\hat\Psi_1 = D\Phi_0, \\
  &(E_2 - H)\hat\Psi_2 = D\hat\Psi_1, \\
  &\qquad\qquad\quad\vdots \nonumber\\
  &(E_K - H)\hat\Psi_K = D\hat\Psi_{K-1},
\end{align}
where $E_j = E_0 + j\omega$, $j = 1,\ldots,K$, and extract the channel
amplitudes from $\hat\Psi_K$. In this case, wave numbers $k_1, \ldots, k_{n-1}$
and $k_n$ correspond to energies $E_1, \ldots, E_{K-1}$ and $E_K$,
respectively. The $K$-photon generalized ionization cross section is then
calculated as \cite{lambropoulos:98}:
\begin{equation}
\sigma^{(K)}_\gamma(\omega) = 2\pi\Big(\frac{2\pi}{c\,\omega}\Big)^K
|\mc{B}_{\gamma}|^2,
\label{eq:sigk}
\end{equation}
where the relation between $\mc{A}_{\gamma,n}$ and $\mc{B}_\gamma$ is given by
Eq.\ \eqref{eq:atob}.
The cross sections can be converted to SI units by multiplying the right-hand
side of Eq.\ \eqref{eq:sigk} with $a_0^{2K}t_0^{K-1}$, where $a_0 \approx
5.29177 \times 10^{-11}$ m is the Bohr radius and $t_0 \approx 2.41888 \times
10^{-17}$ s the atomic unit of time.

\begin{figure}\begin{center}
\includegraphics[width=\linewidth]{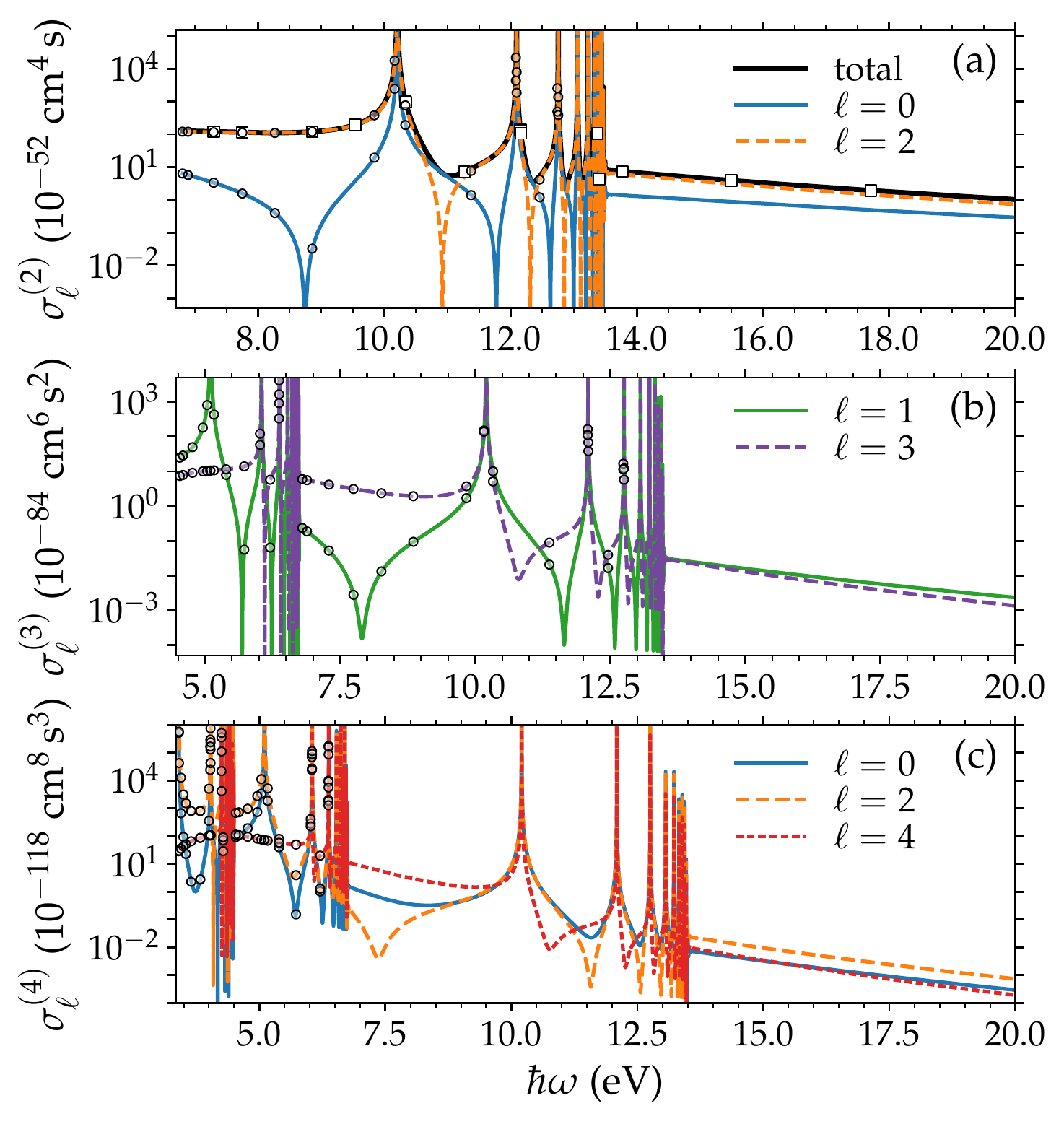}
\caption{Generalized two-, three-, and four-photon partial ionization cross
sections of the hydrogen atom (top to bottom). The partial ionization cross
sections from Ref.\ \cite{karule:88} (circles) and the total two-photon
ionization cross section from Ref.\ \cite{karule:78} (squares) are plotted for
comparison.}
\label{fig:csh}
\end{center}\end{figure}

\section{Results and discussion}

\subsection{One-electron atom}

We used the method described in Section \ref{sec:twoph} to calculate two-photon
ionization cross sections of the ground-state hydrogen atom shown in Fig.\
\ref{fig:csh}(a). The partial and total cross sections (the sum of the $\ell =
0$ and $\ell = 2$ contributions) agree perfectly with the results of the
analytic treatment of Karule \cite{karule:78, karule:88} (circles and squares).
These were calculated from the tabulated values of the intensity-normalized
cross section, $Q^{(2)}_\ell / I \equiv (\hbar\omega)^{-1} \,
\sigma^{(2)}_\ell$, where $I$ is the intensity of the linearly polarized
incident light in W/cm$^2$, $\hbar\omega$ the photon energy, and
$\sigma^{(2)}_\ell$ the generalized cross section in cm$^4\,$s.
Below the ionization threshold ($\hbar\omega < 13.6$ eV), a series of peaks due
to resonance-enhanced ionization through the $np$ states is accompanied by a
series of minima. Strictly speaking, without modifications, the present
formalism is not suitable for photon energies close to the $1s \to np$
transition energies. When the incident photon flux is low, this can be dealt
with by considering the decay and ionization of excited bound states
\cite{lambropoulos:98}. More general alternatives in this case are, for
example, to break up the overall process in steps which may not allow a
perturbative description in terms of transition rates \cite{lambropoulos:98} or
to use the non-perturbative Floquet approach \cite{chu:04}.

In the ATI region ($\hbar\omega > 13.6$ eV), where single-photon ionization of
the atom is possible, the cross sections decrease monotonically with photon
energy.

The three- and four-photon ionization cross sections of the ground-state
hydrogen atom are shown in panels (b) and (c) of Fig.\ \ref{fig:csh}. As can be
seen, our results agree well with the cross sections calculated from the values
of $Q^{(K)}_\ell/I^{K-1} = (\hbar\omega)^{-K+1} \, \sigma^{(K)}_\ell$ tabulated
in Ref.\ \cite{karule:88}. The energy thresholds at approximately 6.8 eV and
13.6 eV in Fig.\ \ref{fig:csh}(b) -- these can be identified by the series of
peaks whose positions converge to these energies -- correspond to the onsets of
the $(2+1)$-photon and $(1+2)$-photon ATI energy regions. Here, $N$ in $N+W$
denotes the number of photons sufficient to ionize the atom and $W$ stands for
the number of photons absorbed above the threshold. Similarly, the thresholds
at approximately 4.5 eV, 6.8 eV, and 13.6 eV mark the start of the $(3+1)$-,
$(2+2)$-, and $(1+3)$-photon ATI energy regions in Fig.\ \ref{fig:csh}(c).

The cross sections shown in Fig.\ \ref{fig:csh} were calculated using a radial
basis of 612 modified $B$-spline functions \cite{mccurdy:04} of order 7 for
each orbital angular momentum $\ell$. The radial integrals were evaluated on
the ECS contour. The radial grid covered an interval up to
$R_{\mathrm{max}} = 250$, and the radial coordinate was scaled ($\vartheta =
0.70$) beyond $R_0 = 200$. Close to the origin ($r = 0$) and for $r \gtrsim
R_0$, a quadratic knot sequence was used; a linear sequence was used elsewhere.
The singular value decomposition (SVD) was used to solve the normal system.

We calculated ionization amplitudes and generalized multiphoton cross sections
up to order $K = 6$, but it should be noted that a calculation of cross
sections of higher orders is possible. The lowest wave number which allows one
to extract the ionization amplitude from $\hat P(r)$ is of the order of
$2\pi/R_0$. In this sense, $R_0$ is one of the critical parameters of the
extraction procedure. Another critical parameter is the number of basis
functions (or, better, the density of collocation points in the radial region
where the amplitudes are extracted); it determines the maximum energy (wave
number) for which an accurate description of the continuum wave functions is
possible \cite{venuti:96, bachau:01}.

\begin{figure*}\begin{center}
\includegraphics[width=\linewidth]{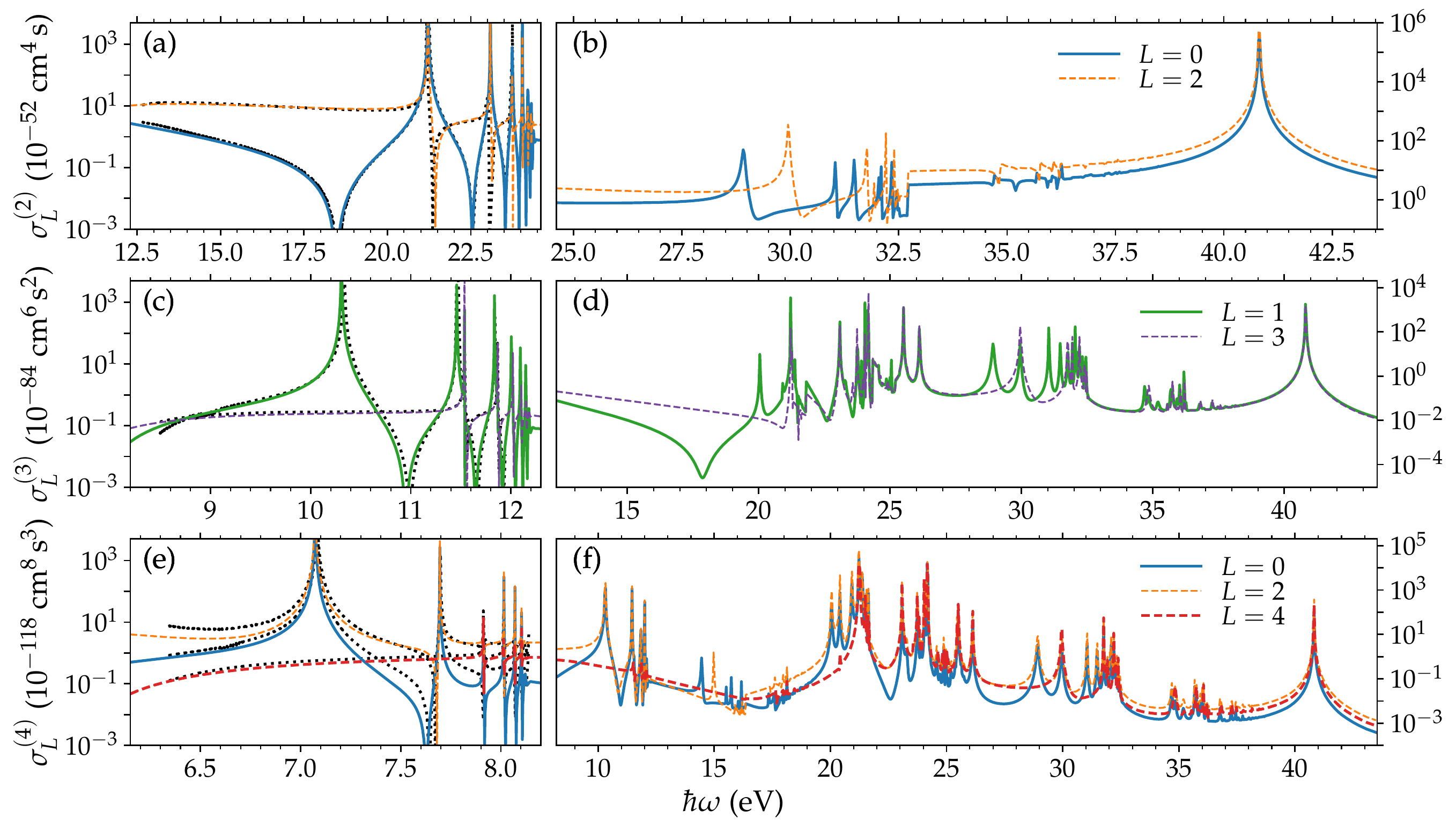}
\caption{Generalized two- (top), three- (middle), and four-photon (bottom)
partial ionization cross sections of the helium atom. The below-threshold
ionization results from Ref.\ \cite{saenz:99} are plotted with dotted lines in
panels (a), (c), and (e).}
\label{fig:cshe}
\end{center}\end{figure*}

\subsection{Two-electron atom}

As noted, the extraction procedure can also be used to calculate multiphoton
ionization amplitudes and cross sections of two- and few-electron atoms. We
tested it on the case of two-, three-, and four-photon ionization of the
ground-state helium atom. In Fig.\ \ref{fig:cshe}, we plot partial ionization
cross sections for linearly polarized ($\uvec{e} = \uvec{z}$) incident light,
for which the final-state channels with $M_L = S = M_S = 0$ are accessible. The
partial cross sections were calculated by summing over all the remaining
channel quantum numbers but the total orbital angular momentum:
\begin{equation}
  \sigma^{(K)}_L(\omega) = \sum_{n_a,\ell_a,\ell}\sigma^{(K)}_\gamma(\omega),
  \label{eq:sigL}
\end{equation}
where $\gamma = (n_a, \ell_a, \ell, L, M_L, S, M_S)$ has been used.  Panels (a),
(c), and (e) show the below-threshold (BTI) energy region. There is generally
good overall agreement between our results and the results of Saenz and
Lambropoulos \cite{saenz:99} (dashed black line), who used the time-independent
perturbation theory to calculate the cross sections. Their continuum states were
calculated by solving the time-independent Schr\"{o}dinger equation subject to
homogeneous boundary conditions in a basis of real $B$-splines \cite{bachau:01}.
Since the photon energy can not be changed independently of the final-state
energy, the scan over the photon energy interval was performed by varying
$R_{\mathrm{max}}$ (see Ref.\ \cite{bachau:01} for details). While two- and
three-photon cross sections from Ref.\ \cite{saenz:99} match the present results
rather well [panels (a) and (c)], slightly larger differences are present in the
case of the four-photon partial cross sections shown in Fig.\ \ref{fig:cshe}(e).
We have checked the validity of the present results by increasing
$R_{\mathrm{max}}$ ($R_0$) and the number of basis functions. Except for the
differences close to the ionization thresholds arising due to additional bound
and resonance states with high principal quantum numbers, the cross sections
remained unchanged. The results in the BTI region were obtained with
single-particle angular momenta up to $\ell_{\mathrm{max}} = 6$ and with the
radial functions for each of the two electrons written in a basis of 150
$B$-splines of order 7 per $\ell$, with $R_{\mathrm{max}} = 170$ and $R_0 =
120$. In the ATI region, we used 85 $B$-splines per $\ell$, $R_{\mathrm{max}} =
85$, and $R_0 = 50$. Partial ionization cross sections of channels with $n_a
\le 5$ were included in the sum [Eq.\ \eqref{eq:sigL}] to obtain the cross
sections in Fig.\ \ref{fig:cshe}. In both energy regions (BTI and ATI),
$\vartheta = 0.70$ was used.

Although not shown, the present results are in excellent agreement with
normalized rates $w^{(3)}_{1s}/I^3 \equiv (\hbar\omega)^{-3} \,
\sigma_{1s}^{(3)}$ reported by Proulx \etal.\ \cite{proulx:94}, where $I$ is
the intensity of the incident light, $\sigma_{1s}^{(3)}$ the sum of partial
three-photon cross sections for ionization leading to the helium ion in the
$1s$ state ($n_a = 1$, $\ell_a = 0$), and $w^{(3)}_{1s}$ the corresponding
ionization rate. These authors used a basis of two-electron Sturmian functions
with complex radial scaling parameters and Pad\'{e} extrapolation to calculate
ionization amplitudes.

A practical note on the implementation of the method is in order. When
some of the energies from the intermediate steps ($E_1, \dots, E_{K-1}$) lie
above the ionization threshold, wave numbers $k_1, \ldots, k_{n-1}$ may be
(approximately) degenerate or may differ by an amount too small to be
``resolved'' when solving the normal system. This is especially true for higher
ionization thresholds. We avoid this by considering only those wave numbers
which differ by more than, say, $\Delta k \sim 10^{-4}$. Furthermore, when
high ionization thresholds are reached (either in the intermediate steps or the
final step), bound states of the atomic core can no longer be adequately
represented in a finite volume ($r \le R_0$). We therefore choose to
additionally limit the wave numbers in the intermediate steps by introducing an
energy cutoff parameter. We have found the results to be stable if
(approximately) equal cutoff energy for the atomic core was used to limit the
intermediate- and final-state ionization channels considered in the
calculation.

In Fig.\ \ref{fig:cshe}, energy thresholds for $(N+W)$-photon ATI may again be
identified by the series of peaks converging to these energies. In the case of
three-photon ionization, for example, the $(2+1)$-photon and $(1+2)$-photon ATI
thresholds lie at 12.3 eV [Fig.\ \ref{fig:cshe}(c)] and 24.6 eV [Fig.\
\ref{fig:cshe}(d)]. In some cases, sharp jumps in $\sigma^{(K)}_L$ mark higher
ionization thresholds, i.e., the thresholds for channels which describe
ionization leading to the ion in an excited state, such as the $2\ell_a
\epsilon_2 \ell$ channels at 32.7 eV, 21.8 eV, and 16.35 eV in panels (b), (d),
and (f) of Fig.\ \ref{fig:cshe}, respectively.
In Fig.\ \ref{fig:cshe}(b), the energy thresholds of the $2\ell_a\epsilon_2\ell$
channels are preceded by a series of peaks with characteristic asymmetric
profiles \cite{sanchez:95} due to the $\LSp{1}{S}{e}$ ($L = 0$) and
$\LSp{1}{D}{e}$ ($L = 2$) autoionizing final states. Their asymmetric shapes
are a consequence of the interference between two ionization pathways: a direct
transition to the $1s\epsilon_2 s$ or $1s\epsilon_2 d$ continuum and an
excitation to a discrete state which is followed by an electron emission.
Asymmetric profiles are also present in the three- and four-photon cross
sections when either the intermediate-state or the final-state continuum (or
both) are resonant.

At $\hbar\omega \approx 40.8$ eV (1.5 a.u.), an enhancement in the cross
sections is a signature of a shake-up like (core-excited) ionization process
\cite{proulx:94, shakeshaft:06, mihelic:18}. This feature is sometimes referred
to as a ``core-excited resonance'' \cite{shakeshaft:06}. The enhancement is a
consequence of a strong laser coupling between the $1s\epsilon_1\ell_1$ and
$2p\epsilon_2\ell_2$ channels. This can be understood if we consider the
following multiphoton ionization pathway:
\begin{align}
    &\text{g.s.} \xrightarrow{\omega} \cdots \xrightarrow{\omega}
    1s\epsilon_1\ell_1 \xrightarrow{\omega} 2p\epsilon_2\ell_2
    \xrightarrow{\omega} \cdots,
    \label{eq:coreex}
\end{align}
where g.s.\ stands for the helium ground state. In Eq.\ \eqref{eq:coreex}, the
step written symbolically as $1s\epsilon_1\ell_1 \to 2p\epsilon_2\ell_2$
describes a CC transition, for which Eq.\ \eqref{eq:onshell} may be used.  We
thus expect that the $1s\epsilon_1\ell_1$ ionization channel is strongly
coupled to the $2p\epsilon_2\ell_2$ channel when $\epsilon_2\ell_2 =
\epsilon_1\ell_1$, i.e., when the continuum electron acts as a spectator and
the electronic transition happens in the atomic core \cite{mihelic:18,
  proulx:94}. This is possible when the photon energy is approximately equal to
the transition energy of the core ($\hbar\omega \approx I_{2p} - I_{1s}$). For
example, in the case of two-photon ionization [Fig.\ \ref{fig:cshe}(b)], the
peak at 40.8 eV is due to the laser coupling between the $1s\epsilon_1 p$ and
$1s\epsilon_2 p$ ionization channels ($\text{g.s.} \to 1s\epsilon_1p \to
2p\epsilon_2 p$). Note that although the photon energy of 40.8 eV coincides
with the $1s \to 2p$ transition energy in He$^+$, Eq.\ \eqref{eq:coreex}
describes a multiphoton process in a neutral atom, i.e., it does not correspond
to a two-part (``sequential'') process, in which the atom is first ionized
($\text{He} + \gamma \to \text{He}^+\,1s + e^-$) and the ground-state ion is
then excited to a higher-lying state ($\text{He}^+\,1s + \gamma \to 2p$)
\cite{proulx:94}. Similar enhancements also occur at photon energies
$\hbar\omega = I_{n_a} - I_{1s}$ (48.4 eV, 51.0 eV, 52.2 eV, etc.)
\cite{shakeshaft:06, mihelic:18}, where $n_a$ denotes the principal quantum
number of the excited state of the atomic core. These peaks are associated with
the $1s\epsilon_1\ell \to n_ap\epsilon_2\ell$ CC transitions.

As has been discussed in Ref.\ \cite{mihelic:18}, the structure of the
differential equation for the radial part of the driven Schr\"{o}dinger
equation describing the $2p\epsilon_2\ell_2$ channels resembles the equation of
motion of a driven harmonic oscillator in which $r$ (instead of time $t$) is
the independent variable, and the frequency of the oscillator and the driving
frequency are replaced by $k_2$ and $k_1$, respectively. When $\hbar\omega =
I_{2p} - I_{1s}$ ($k_1 = k_2$), the situation corresponds to a resonantly
driven, nondamped harmonic oscillator. Exactly on resonance, the present method
will fail \cite{mihelic:18}. The reason for this is that no spontaneous decay
(of the core vacancy) or field-dressing effects have been included in the
present formalism \cite{shakeshaft:06}. As in the case of bound intermediate
states, this may be addressed in the framework of the (non-perturbative)
Floquet formalism \cite{chu:04}.

\begin{figure}\begin{center}
\includegraphics[width=\linewidth]{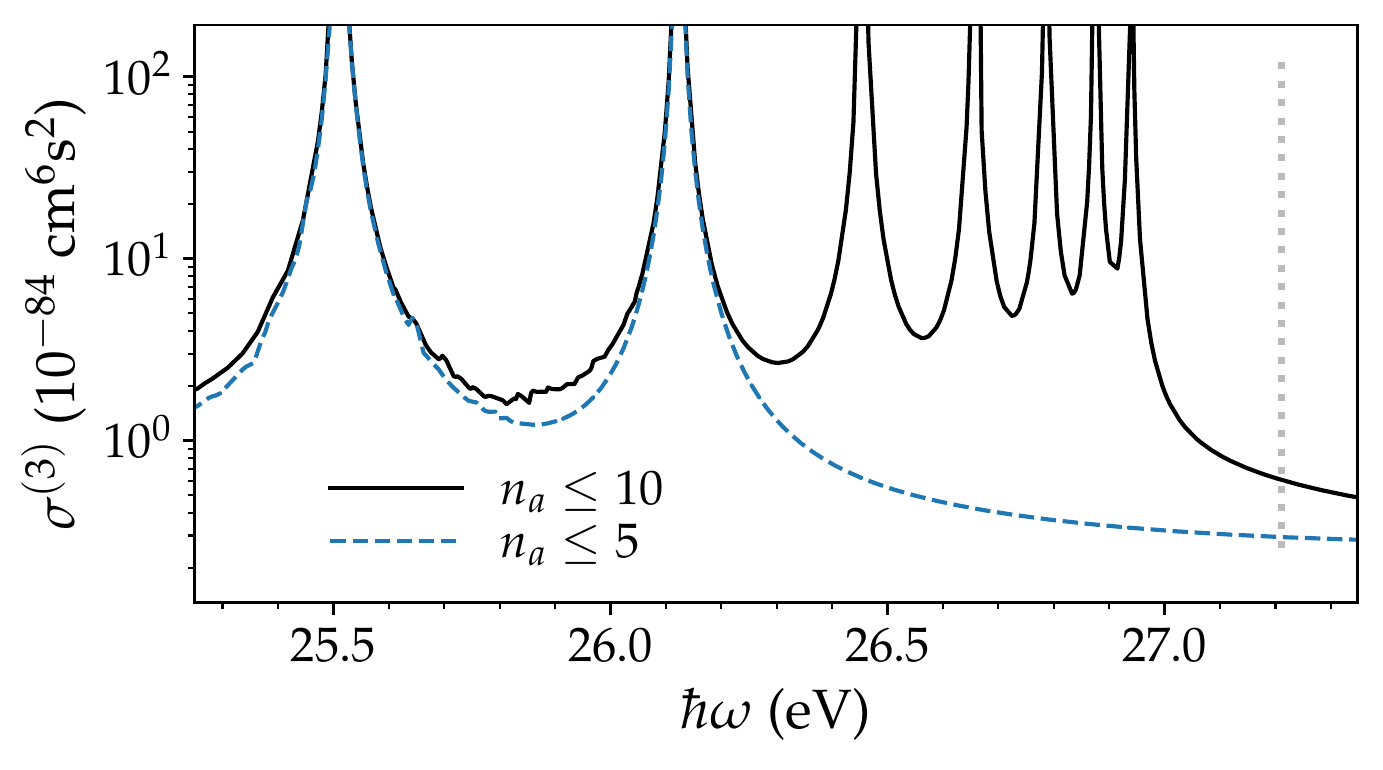}
\caption{Three-photon ionization cross section in the region of $(1+2)$-photon
ATI. Results of two separate calculations are shown in which ionization
channels with $n_a \le 5$ (dashed blue line) and $n_a \le 10$ (solid black
line) have been included.}
\label{fig:cs3core}
\end{center}\end{figure}

Photoionization which is accompanied by core excitation can also be observed at
lower photon energies. One such example is $(1+2)$-photon ATI for photon
energies between 25.2 eV and 27.2 eV [Fig.\ \ref{fig:cshe}(d)]. An enlarged
view of this energy interval is plotted in Fig.\ \ref{fig:cs3core}. The total
three-photon cross section from two separate calculations is shown: in the
first, channels with $n_a \le 5$ have been included; in the second, $R_0 =
120$, $R_{\mathrm{max}} = 160$, and a larger basis set (120 $B$-splines per
$\ell$ for each electron) were used to calculate partial ionization cross
sections for channels with $n_a \le 10$. As can be seen, the peak positions
converge to the limit of 1 a.u. (27.2 eV), which is marked with a dotted
vertical line in Fig.\ \ref{fig:cs3core}. The dominant ionization pathways
underlying these enhancements are of the form $\text{g.s.} \to 1s\epsilon_1 p
\to 1s\epsilon_2\ell \to n_a p\epsilon_3\ell$, with $n_a \ge 4$ and $\ell =
s, d$.

\begin{figure}\begin{center}
\includegraphics[width=\linewidth]{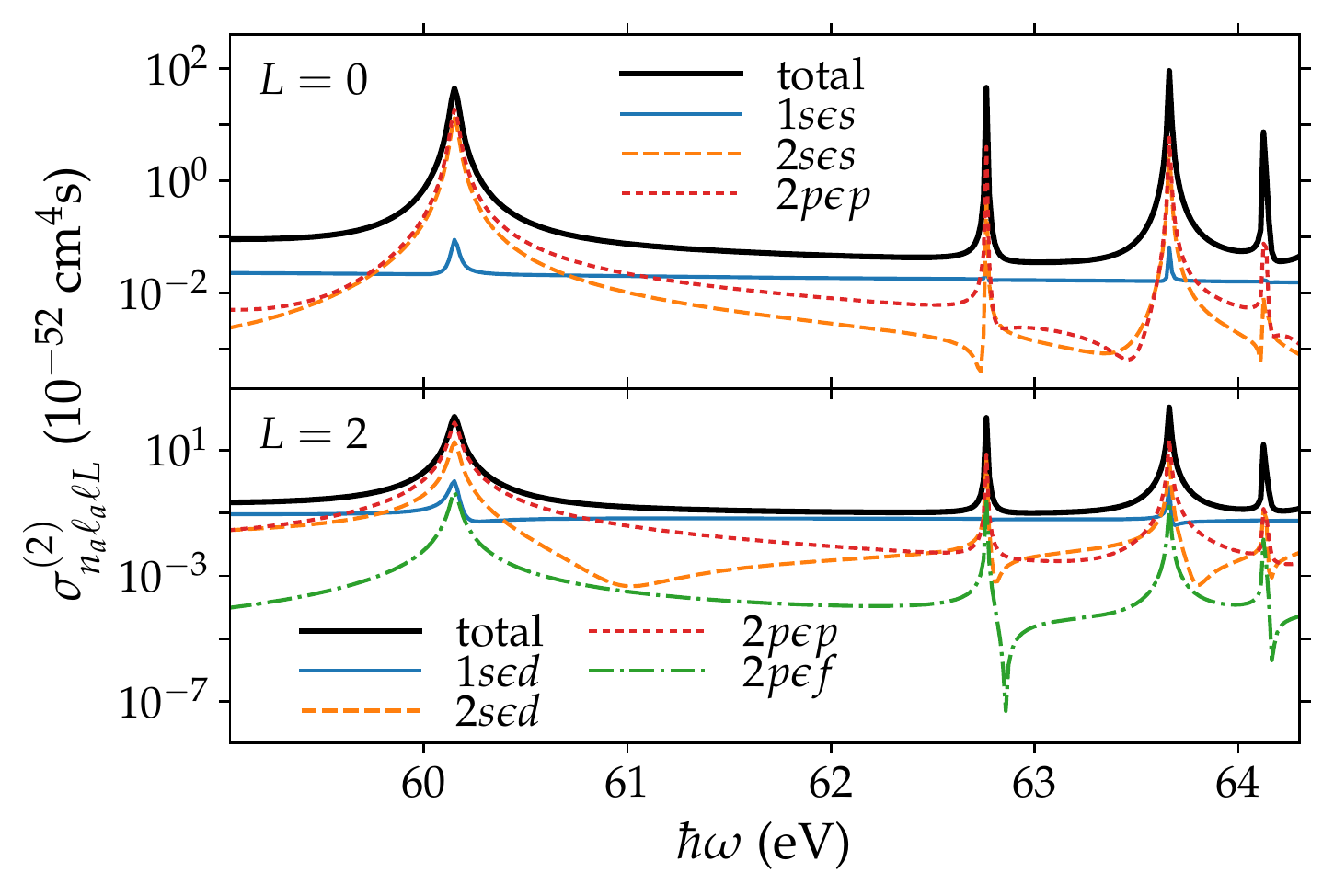}
\caption{Partial two-photon generalized cross sections for the $n_a\ell_a
\epsilon\ell \, L$ ionization channels in the energy region of the lowest
intermediate $\LSp{1}{P}{o}$ autoionizing states: $sp^+_2$, $sp^-_3$, $sp^+_3$,
and $2p3d$ (corresponding to peaks with increasing energy).}
\label{fig:rempi}
\end{center}\end{figure}

The present method allows one to calculate partial ionization amplitudes and
cross sections also in the case of resonance-enhanced multiphoton ionization
(REMPI), i.e., when the driving is resonant with a bound or a quasi-bound
(resonance) state in the intermediate step. In Fig.\ \ref{fig:rempi}, we show
partial two-photon ionization cross sections in the energy region of the lowest
$\LSp{1}{P}{o}$ autoionizing (doubly excited) resonance states: $sp^+_2$,
$sp^-_3$, $sp^+_3$, and $2p3d$. We have used the notation of Cooper, Fano,
and Pratts \cite{cooper:63}. These autoionizing states, which lie below the
second ionization threshold (65.4 eV), can be described primarily with the
$2s2p$, $2p3s$, $2s3p$, and $2p3d$ configuration basis states. As seen in Fig.\
\ref{fig:rempi}, as well as in Figs.\ \ref{fig:csh} and \ref{fig:cshe}, the
method introduces no additional broadening, either below or above the
ionization threshold.

\subsection{Photoelectron angular distributions}

Let us conclude the discussion by noting that it is possible to calculate
photoelectron angular distributions (PADs) of helium from channel amplitudes
$\mc{B}_\gamma$ \cite{okeeffe:10, okeeffe:13, carette:13}, where, as before,
$\gamma = (n_a, \ell_a, \ell, L, M_L, S, M_S)$ has been used as a short-hand
notation for the quantum numbers of the ionization channel. We write the
angle-dependent photoelectron amplitude as:
\begin{align}
  \mc{G}_a(\vec{k}) =
  \sum_{L,\ell,m}\cg{\ell_a}{,m_a}{;\ell}{,m}{L}{M_L} \mc{B}_\gamma
  Y_{\ell{}m}^*(\uvec{k}),
  \label{eq:angleamp}
\end{align}
where $\vec{k}$ denotes the wave vector of the ejected electron, $\uvec{k} =
\vec{k}/k$, $m_a$ is the projection of the orbital angular momentum of the core
and $\cg{\ell_a} {,m_a} {;\ell} {,m} {L}{M_L}$ is the Clebsch-Gordan (vector
coupling) coefficient \cite{brink:75}. The $K$-photon differential cross
section is then proportional to:
\begin{align}
  \frac{\dd\sigma^{(K)}}{\dd\Omega_{\vec{k}}}
  \propto \sum_{n_a,\ell_a,m_a} |\mc{G}_a(\vec{k})|^2.
  \label{eq:diffcs}
\end{align}
By omitting the Clebsch-Gordan coefficient for the spin in Eq.\
\eqref{eq:angleamp}, we
have implicitly summed (``averaged'') over the spin quantum numbers of the
target. In the present case (linearly polarized light, $\uvec{e} = \uvec{z}$),
only partial waves with $M_L = 0$ are accessible; the PADs will be axially
symmetric in this case. By using the addition theorem for the spherical
harmonics and the reduction formula for the $3j$ symbols \cite{brink:75}, Eq.\
\eqref{eq:diffcs} simplifies to:
\begin{align}
  \sum_{n_a,\ell_a,m_a} |\mc{G}_a(\vec{k})|^2
  = \sum_{j = 0}^{K} \mc{N}_{2j} P_{2j}(\uvec{e}\cdot\uvec{k}),
  \label{eq:legendre}
\end{align}
where $P_\kappa(\uvec{e} \cdot \uvec{k}) = \sqrt{4\pi/(2\kappa+1)}
Y_{\kappa,0}(\uvec{k})$ is the Legendre polynomial of order $\kappa$. The
coefficients in Eq.\ \eqref{eq:legendre} are:
\begin{equation}
\begin{aligned}
  \mc{N}_\kappa = \sum_{n_a,\ell_a} \sum_{L,L'}\sum_{\ell,\ell'}
      (-1)^{\kappa+\ell_a+\ell+\ell'}
  \widehat{\kappa}^2
  \widehat{\ell}
  \widehat{\ell'}
  \widehat{L}
  \widehat{L'} \, \times \\
  \threej{\ell}{\ell'}{\kappa}{0}{0}{0}
  \threej{L}{L'}{\kappa}{0}{0}{0}
  \sixj{L}{L'}{\kappa}{\ell'}{\ell}{\ell_a}
  \mc{B}_{\gamma'}\mc{B}^*_\gamma.
  \label{eq:nk}
\end{aligned}
\end{equation}
A shorthand notation $\widehat{a} = \sqrt{2a + 1}$ has been introduced, and it
is to be understood that indices $\gamma$ and $\gamma'$ are used in place of
$(n_a,\ell_a,\ell,L)$ and $(n_a,\ell_a,\ell',L')$. We may now define the
asymmetry parameters as:
\begin{equation}
  \beta_\kappa = \mc{N}_\kappa/\mc{N}_0,
  \label{eq:beta}
\end{equation}
where the expression for $\mc{N}_0$ reduces to
\begin{equation}
  \mc{N}_0 = \sum_{n_a,\ell_a} \sum_{L,\ell} |\mc{B}_\gamma|^2
  \label{eq:n0}
\end{equation}
after evaluating the $3j$ and $6j$ symbols for $\ell = \ell'$ and $L = L'$.
Equations \eqref{eq:nk}--\eqref{eq:n0} agree with the analogous expressions
given in Ref.\ \cite{boll:19}.

\begin{figure}\begin{center}
  \includegraphics[width=\linewidth]{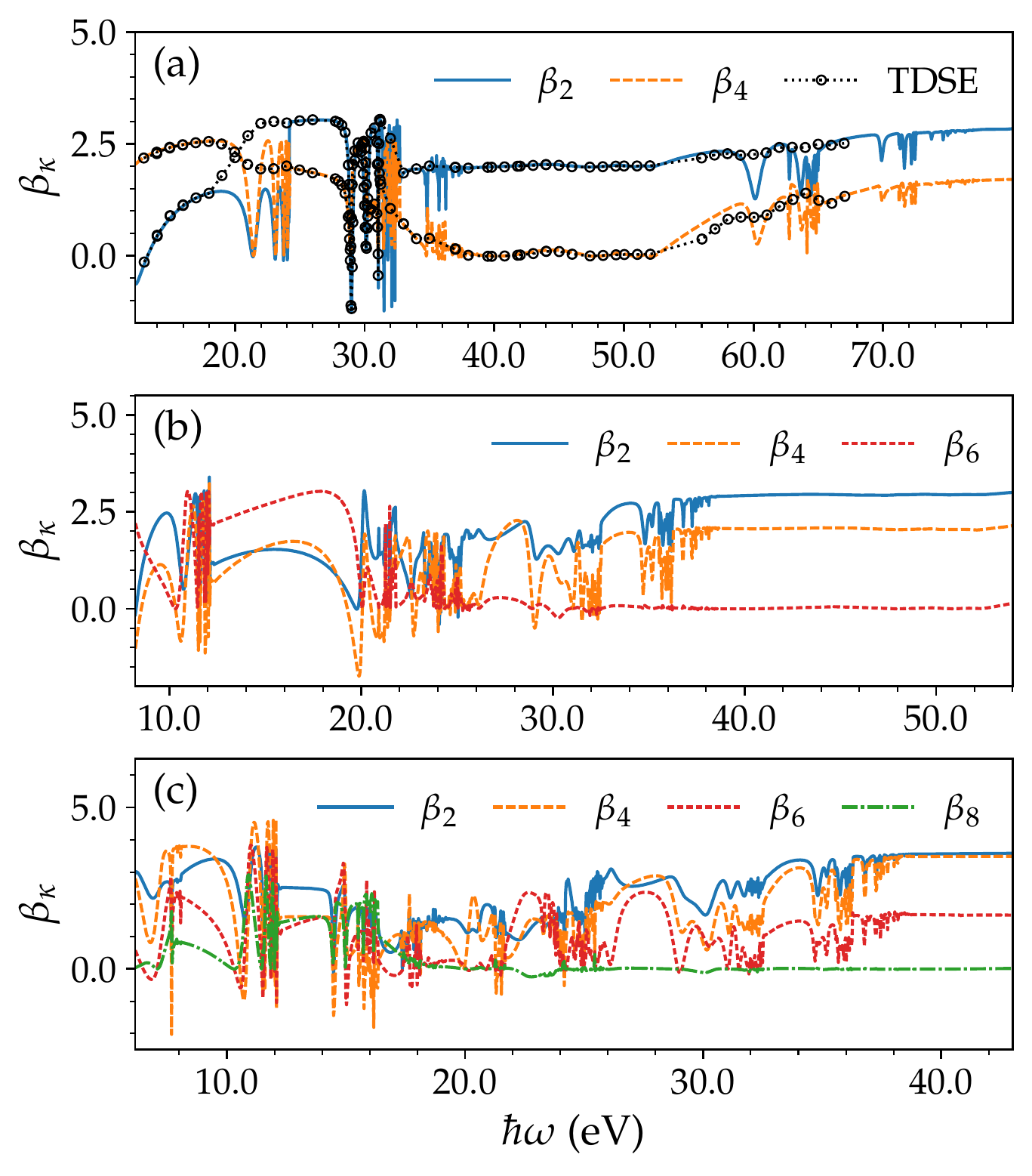}
  \caption{Asymmetry parameters $\beta_\kappa$ for two-, three-, and
  four-photon ionization of the He atom (top to bottom). The results from the
  time-dependent calculation (TDSE) in panel (a) are from Ref.\
  \cite{boll:19}.}
\label{fig:beta}
\end{center}\end{figure}

In Fig.\ \ref{fig:beta}, we show the asymmetry parameters for two-, three-, and
four-photon ionization of He.
The strong variation of $\beta_2$ and $\beta_4$ in the energy region below the
first ionization threshold (24.6 eV) seen in Fig.\ \ref{fig:beta}(a) is a
consequence of two-photon ionization through the (bound) $1snp$ states. For
photon energies between 30 and 40 eV, autoionizing states whose
energies converge to the second ionization threshold (65.4 eV), as well as
higher thresholds, can be reached; the resonant nature of the final continuum
states results in a strong energy dependence of the asymmetry parameters. The
parameters remain almost constant in the energy region where core-excited
ionization is dominant (roughly between 40.8 eV and 54.4 eV). Finally, at even higher
energies, for $\hbar\omega \gtrsim 60$ eV, the changes in the asymmetry
parameters are due to two-photon ionization which proceeds through the
odd-parity autoionizing states (REMPI), as shown in Fig.\ \ref{fig:rempi} and
discussed above.
The calculated asymmetry parameters are in good agreement with the results of
the recent time-dependent calculation of Boll \etal.\ \cite{boll:19}, in
which channel amplitudes have been extracted from the wave packet at the end of
the laser pulse. Note that the main differences stem from the energy broadening
due to the finite duration of the laser pulse (a 2 fs long pulse was used in
Ref.\ \cite{boll:19}).

It may come as a surprise that the asymmetry parameters in
Fig.\ \ref{fig:beta}(a) vary weakly with photon energy in the region of
core-excited ionization. It has been checked that the phase shifts of the
dominant $2p\epsilon_2 p$ channels are indeed smooth in this energy region.
Furthermore, the values of the asymmetry parameters in this region are very
close to $\beta_2 = 2$ and $\beta_4 = 0$, which describe the PADs in the case
of single-photon ionization for photon energies below the second ionization
threshold. This confirms that the enhancement of the cross sections is a
consequence of an ``external'' interaction (interaction with the laser), which
causes a transition inside the atomic core, but does not alter the (phases of
the) final-state partial waves. Conversely, the phase shifts of the
$1s\epsilon_2\ell$ ($\ell = s, d$), as well as the $2s \epsilon_2 \ell$ and $2p
\epsilon_2 f$ channels all exhibit abrupt jumps at 40.8 eV. The latter are
indicators of the coupling of these channels with the optically-accessible
$2p\epsilon_2 p$ channels, i.e., the jumps are a direct consequence of the
electron-electron (Coulomb) coupling in the final state. Note however, that the
cross sections corresponding to these channels are several orders of magnitude
lower than the $2p\epsilon_2 p$ cross sections.

An analogous behavior may also be seen for the asymmetry parameters in the case
of three-photon [Fig.\ \ref{fig:beta}(b)] and four-photon [Fig.\
\ref{fig:beta}(c)] ionization. The parameters vary strongly when the photon
energy is close to intermediate bound or resonance states or when the
final-state continuum is resonant. In the region of core-excited ionization,
the asymmetry parameters again vary smoothly with photon energy.

\section{Conclusions}

We have devised a theoretical method for the calculation of multiphoton
ionization amplitudes and cross sections of few-electron atoms. The method is
based on extraction of partial wave amplitudes from the scattered part of a wave
function, which is obtained by solving a set of driven Schr\"{o}dinger
equations, and works in the case of a single electron ejection. The extraction
procedure relies on a description of partial waves in terms of a small number
of Coulomb waves with fixed wave numbers.
We have implemented the extraction procedure in the framework of exterior
complex scaling. One- and two-electron wave functions have been calculated in a
basis of $B$-spline functions.
The method has been tested by calculating partial two-, three-, and four-photon
generalized cross sections of atomic hydrogen and helium, and by calculating
the asymmetry parameters of photoelectron angular distributions for two-,
thee-, and four-photon ionization of the helium atom.

We have found the present method to be stable, robust, and its implementation
to be relatively straightforward. While the method has only been tested on
atomic systems, it could also be used to treat multiphoton
ionization of simple molecules.

\begin{acknowledgments}

We acknowledge the financial support from the Slovenian Research Agency
(research programs No.\ P1--0112 and No.\ P1--0402 and research projects No.\ J1--8134 and No.\
J1--1698). This work was supported by the European COST Action CA$\,$18222
(AttoChem). The calculations were carried out on the Olimp computer cluster at
the Faculty of mathematics and physics. We thank Blaž Jesenko for his support
and smooth operation of the computational facilities.

\end{acknowledgments}

\appendix

\section{Calculation of radial function $\hat P(r)$}
\label{sec:calcp}

In this section, we provide additional details on the calculation of
photoionization amplitudes for a two-electron atom. In particular, we discuss
one of the possible ways of calculating radial function $\hat P(r)$ associated
with ionization channel $\gamma = (n_a, \ell_a, \ell, L, M_L, S, M_S)$ of
$\hat\Psi(\vec{r}_1, \vec{r}_2)$, the solution of Eq.\ \eqref{eq:psisc}. We
start by expressing $\hat\Psi(\vec{r}_1, \vec{r}_2)$ as:
\begin{equation}
   \hat\Psi(\vec{r}_1, \vec{r}_2) = \sum_{L',M_{L'}}
   \sum_{\alpha,\beta} y^{L'M_{L'}}_{\alpha\beta}
       \Phi^{L'M_{L'}}_{\alpha\beta}(\vec{r}_1,\vec{r}_2),
   \label{eq:expa}
\end{equation}
where $\Phi^{L'M_{L'}}_{\alpha\beta}(\vec{r}_1,\vec{r}_2)$ denotes a coupled
two-electron wave function:
\begin{equation}
   \Phi^{L'M_{L'}}_{\alpha\beta}(\vec{r}_1,\vec{r}_2) =
   \mf{A} \frac{P_{n_\alpha\ell_\alpha}(r_1)}{r_1}
   \frac{P_{n_\beta\ell_\beta}(r_2)}{r_2}
   \mc{Y}^{\ell_\alpha\ell_\beta}_{L'M_{L'}}(\uvec{r}_1,\uvec{r}_2).
   \label{eq:phiab}
\end{equation}
In Eq.\ \eqref{eq:phiab}, $\mc{Y}^{\ell_\alpha\ell_\beta}_{L'M_{L'}}
(\uvec{r}_1, \uvec{r}_2)$ is a bipolar harmonic \cite{brink:75} and $\mf{A}$
stands for the antisymmetrizing operator. (The spin parts have been omitted for
brevity.) Radial functions $P_{n_\alpha\ell_\alpha}(r)$ are obtained by
calculating eigen wave functions of the complex-scaled Hamiltonian operator of
a one-electron atom with nuclear charge $Z$. These functions can be used to
represent both bound and continuum states.  Next, we calculate the projection
by fixing the quantum numbers of the atomic core ($n_\alpha = n_a$,
$\ell_\alpha = \ell_a$), the total angular momentum and its projection ($L' =
L$, $M_{L'} = M_L$), and the orbital angular momentum of the ``outer'' electron
($\ell_\beta = \ell$):
\begin{equation}
   \hat\Psi^{LM_L}_{a\ell}(\vec{r}_1, \vec{r}_2) = \sum_\beta
   \Phi^{LM_L}_{a\beta}(\vec{r}_1, \vec{r}_2)
   \braket{\Phi^{LM_L}_{a\beta}}{\hat\Psi}\delta_{\ell,\ell_\beta},
   \label{eq:psiproj}
\end{equation}
where the overlap matrix element $\braket{\Phi^{LM_L}_{a\beta}}{\hat\Psi}$ is
calculated on the ECS contour. It has been assumed above that $P_{n_a\ell_a}(r)$
describes a bound state whose wave function is contained within the nonscaled
radial region, i.e., $P_{n_a\ell_a}(r)$ is taken to be negligibly small for $r >
R_0$. Since wave functions $\Phi^{LM_L}_{\alpha \beta}(\vec{r}_1, \vec{r}_2)$
are diagonal in all quantum numbers, Eq.\ \eqref{eq:psiproj} may be written as:
\begin{equation}
   \hat\Psi^{LM_L}_{a\ell}(\vec{r}_1, \vec{r}_2) = \sum_\beta
   \Phi^{LM_L}_{a\beta}(\vec{r}_1, \vec{r}_2)
   y^{LM_L}_{a\beta} \delta_{\ell,\ell_\beta}.
   \label{eq:psiproj2}
\end{equation}
We may now immediately write the radial function describing the continuum
electron in channel $\gamma$ as:
\begin{equation}
   \hat P(r) = \sum_\beta y^{LM_L}_{a\beta}
   P_{n_\beta\ell_\beta}(r) \delta_{\ell,\ell_\beta}.
\end{equation}

Coefficients $y^{LM_L}_{a\beta}$ are calculated in the following way. We write
$\hat\Psi$ and $\Phi^{LM_L}_{\alpha\beta}$ in a basis of two-electron functions
\begin{equation}
   \varphi_i(\vec{r}_1,\vec{r}_2) = \mf{A}
   \frac{B_{n_i}(r_1)}{r_1}
   \frac{B_{\nu_i}(r_2)}{r_2}
   \mc{Y}^{\ell_i\lambda_i}_{L_iM_{L_i}}(\uvec{r}_1,\uvec{r}_2),
\end{equation}
where $B_{n_i}$ and $B_{\nu_i}$ are the modified $B$-spline functions
\cite{mccurdy:04}, $\ell_i$ and $\lambda_i$ denote the orbital angular momenta
of the two electrons, and $L_i$ and $M_i$ the total orbital angular momentum and
its projection. Coefficients $y^{LM_L}_{a\beta}$ are then expressed as:
\begin{align}
  y^{LM_L}_{a\beta} = \sum_{i,j} v^{a\beta}_i \mc{S}_{ij} w_j,
  \label{eq:yab}
\end{align}
where $\mc{S}_{ij} = \braket{\varphi_i}{\varphi_j}$ is the overlap matrix
element evaluated on the ECS contour, and $w_j$ and $v^{\alpha\beta}_i$ denote
the expansion coefficients of $\hat\Psi$ and $\Phi^{LM_L}_{\alpha\beta}$:
\begin{align}
  \hat\Psi(\vec{r}_1,\vec{r}_2) &=
      \sum_j w_j \varphi_j(\vec{r}_1,\vec{r}_2), \\
  \Phi^{LM_L}_{\alpha\beta}(\vec{r}_1,\vec{r}_2) &=
      \sum_i v^{\alpha\beta}_i \varphi_i(\vec{r}_1,\vec{r}_2).
\end{align}

\bibliographystyle{apsrev4-1}
\bibliography{main}

\begin{thebibliography}{43}%
\makeatletter
\providecommand \@ifxundefined [1]{%
 \@ifx{#1\undefined}
}%
\providecommand \@ifnum [1]{%
 \ifnum #1\expandafter \@firstoftwo
 \else \expandafter \@secondoftwo
 \fi
}%
\providecommand \@ifx [1]{%
 \ifx #1\expandafter \@firstoftwo
 \else \expandafter \@secondoftwo
 \fi
}%
\providecommand \natexlab [1]{#1}%
\providecommand \enquote  [1]{``#1''}%
\providecommand \bibnamefont  [1]{#1}%
\providecommand \bibfnamefont [1]{#1}%
\providecommand \citenamefont [1]{#1}%
\providecommand \href@noop [0]{\@secondoftwo}%
\providecommand \href [0]{\begingroup \@sanitize@url \@href}%
\providecommand \@href[1]{\@@startlink{#1}\@@href}%
\providecommand \@@href[1]{\endgroup#1\@@endlink}%
\providecommand \@sanitize@url [0]{\catcode `\\12\catcode `\$12\catcode
  `\&12\catcode `\#12\catcode `\^12\catcode `\_12\catcode `\%12\relax}%
\providecommand \@@startlink[1]{}%
\providecommand \@@endlink[0]{}%
\providecommand \url  [0]{\begingroup\@sanitize@url \@url }%
\providecommand \@url [1]{\endgroup\@href {#1}{\urlprefix }}%
\providecommand \urlprefix  [0]{URL }%
\providecommand \Eprint [0]{\href }%
\providecommand \doibase [0]{http://dx.doi.org/}%
\providecommand \selectlanguage [0]{\@gobble}%
\providecommand \bibinfo  [0]{\@secondoftwo}%
\providecommand \bibfield  [0]{\@secondoftwo}%
\providecommand \translation [1]{[#1]}%
\providecommand \BibitemOpen [0]{}%
\providecommand \bibitemStop [0]{}%
\providecommand \bibitemNoStop [0]{.\EOS\space}%
\providecommand \EOS [0]{\spacefactor3000\relax}%
\providecommand \BibitemShut  [1]{\csname bibitem#1\endcsname}%
\let\auto@bib@innerbib\@empty
\bibitem [{\citenamefont {Mainfray}\ and\ \citenamefont
  {Manus}(1991)}]{mainfray:91}%
  \BibitemOpen
  \bibfield  {author} {\bibinfo {author} {\bibfnamefont {G.}~\bibnamefont
  {Mainfray}}\ and\ \bibinfo {author} {\bibfnamefont {G.}~\bibnamefont
  {Manus}},\ }\href {\doibase 10.1088/0034-4885/54/10/002} {\bibfield
  {journal} {\bibinfo  {journal} {Rep. Prog. Phys.}\ }\textbf {\bibinfo
  {volume} {54}},\ \bibinfo {pages} {1333} (\bibinfo {year}
  {1991})}\BibitemShut {NoStop}%
\bibitem [{\citenamefont {Chin}\ and\ \citenamefont
  {Lambropoulos}(1984)}]{chin:84}%
  \BibitemOpen
  \bibinfo {editor} {\bibfnamefont {S.~L.}\ \bibnamefont {Chin}}\ and\ \bibinfo
  {editor} {\bibfnamefont {P.}~\bibnamefont {Lambropoulos}},\ eds.,\ \href@noop
  {} {\emph {\bibinfo {title} {Multiphoton ionization of atoms}}}\ (\bibinfo
  {publisher} {Academic Press},\ \bibinfo {address} {Toronto},\ \bibinfo {year}
  {1984})\BibitemShut {NoStop}%
\bibitem [{\citenamefont {Ott}\ \emph {et~al.}(2014)\citenamefont {Ott},
  \citenamefont {Kaldun}, \citenamefont {Argenti}, \citenamefont {Raith},
  \citenamefont {Meyer}, \citenamefont {Laux}, \citenamefont {Zhang},
  \citenamefont {Blättermann}, \citenamefont {Hagstotz}, \citenamefont {Ding},
  \citenamefont {Heck}, \citenamefont {Madroñero}, \citenamefont {Martín},\
  and\ \citenamefont {Pfeifer}}]{ott:14}%
  \BibitemOpen
  \bibfield  {author} {\bibinfo {author} {\bibfnamefont {C.}~\bibnamefont
  {Ott}}, \bibinfo {author} {\bibfnamefont {A.}~\bibnamefont {Kaldun}},
  \bibinfo {author} {\bibfnamefont {L.}~\bibnamefont {Argenti}}, \bibinfo
  {author} {\bibfnamefont {P.}~\bibnamefont {Raith}}, \bibinfo {author}
  {\bibfnamefont {K.}~\bibnamefont {Meyer}}, \bibinfo {author} {\bibfnamefont
  {M.}~\bibnamefont {Laux}}, \bibinfo {author} {\bibfnamefont {Y.}~\bibnamefont
  {Zhang}}, \bibinfo {author} {\bibfnamefont {A.}~\bibnamefont {Blättermann}},
  \bibinfo {author} {\bibfnamefont {S.}~\bibnamefont {Hagstotz}}, \bibinfo
  {author} {\bibfnamefont {T.}~\bibnamefont {Ding}}, \bibinfo {author}
  {\bibfnamefont {R.}~\bibnamefont {Heck}}, \bibinfo {author} {\bibfnamefont
  {J.}~\bibnamefont {Madroñero}}, \bibinfo {author} {\bibfnamefont
  {F.}~\bibnamefont {Martín}}, \ and\ \bibinfo {author} {\bibfnamefont
  {T.}~\bibnamefont {Pfeifer}},\ }\href {\doibase 10.1038/nature14026}
  {\bibfield  {journal} {\bibinfo  {journal} {Nature}\ }\textbf {\bibinfo
  {volume} {516}},\ \bibinfo {pages} {374} (\bibinfo {year}
  {2014})}\BibitemShut {NoStop}%
\bibitem [{\citenamefont {Prince}\ \emph {et~al.}(2016)\citenamefont {Prince},
  \citenamefont {Allaria}, \citenamefont {Callegari}, \citenamefont {Cucini},
  \citenamefont {De~Ninno}, \citenamefont {Di~Mitri}, \citenamefont {Diviacco},
  \citenamefont {Ferrari}, \citenamefont {Finetti}, \citenamefont {Gauthier},
  \citenamefont {Giannessi}, \citenamefont {Mahne}, \citenamefont {Penco},
  \citenamefont {Plekan}, \citenamefont {Raimondi}, \citenamefont {Rebernik},
  \citenamefont {Roussel}, \citenamefont {Svetina}, \citenamefont {Trovò},
  \citenamefont {Zangrando}, \citenamefont {Negro}, \citenamefont
  {Carpeggiani}, \citenamefont {Reduzzi}, \citenamefont {Sansone},
  \citenamefont {Grum-Grzhimailo}, \citenamefont {Gryzlova}, \citenamefont
  {Strakhova}, \citenamefont {Bartschat}, \citenamefont {Douguet},
  \citenamefont {Venzke}, \citenamefont {Iablonskyi}, \citenamefont {Kumagai},
  \citenamefont {Takanashi}, \citenamefont {Ueda}, \citenamefont {Fischer},
  \citenamefont {Coreno}, \citenamefont {Stienkemeier}, \citenamefont
  {Ovcharenko}, \citenamefont {Mazza},\ and\ \citenamefont
  {Meyer}}]{prince:16}%
  \BibitemOpen
  \bibfield  {author} {\bibinfo {author} {\bibfnamefont {K.~C.}\ \bibnamefont
  {Prince}}, \bibinfo {author} {\bibfnamefont {E.}~\bibnamefont {Allaria}},
  \bibinfo {author} {\bibfnamefont {C.}~\bibnamefont {Callegari}}, \bibinfo
  {author} {\bibfnamefont {R.}~\bibnamefont {Cucini}}, \bibinfo {author}
  {\bibfnamefont {G.}~\bibnamefont {De~Ninno}}, \bibinfo {author}
  {\bibfnamefont {S.}~\bibnamefont {Di~Mitri}}, \bibinfo {author}
  {\bibfnamefont {B.}~\bibnamefont {Diviacco}}, \bibinfo {author}
  {\bibfnamefont {E.}~\bibnamefont {Ferrari}}, \bibinfo {author} {\bibfnamefont
  {P.}~\bibnamefont {Finetti}}, \bibinfo {author} {\bibfnamefont
  {D.}~\bibnamefont {Gauthier}}, \bibinfo {author} {\bibfnamefont
  {L.}~\bibnamefont {Giannessi}}, \bibinfo {author} {\bibfnamefont
  {N.}~\bibnamefont {Mahne}}, \bibinfo {author} {\bibfnamefont
  {G.}~\bibnamefont {Penco}}, \bibinfo {author} {\bibfnamefont
  {O.}~\bibnamefont {Plekan}}, \bibinfo {author} {\bibfnamefont
  {L.}~\bibnamefont {Raimondi}}, \bibinfo {author} {\bibfnamefont
  {P.}~\bibnamefont {Rebernik}}, \bibinfo {author} {\bibfnamefont
  {E.}~\bibnamefont {Roussel}}, \bibinfo {author} {\bibfnamefont
  {C.}~\bibnamefont {Svetina}}, \bibinfo {author} {\bibfnamefont
  {M.}~\bibnamefont {Trovò}}, \bibinfo {author} {\bibfnamefont
  {M.}~\bibnamefont {Zangrando}}, \bibinfo {author} {\bibfnamefont
  {M.}~\bibnamefont {Negro}}, \bibinfo {author} {\bibfnamefont
  {P.}~\bibnamefont {Carpeggiani}}, \bibinfo {author} {\bibfnamefont
  {M.}~\bibnamefont {Reduzzi}}, \bibinfo {author} {\bibfnamefont
  {G.}~\bibnamefont {Sansone}}, \bibinfo {author} {\bibfnamefont {A.~N.}\
  \bibnamefont {Grum-Grzhimailo}}, \bibinfo {author} {\bibfnamefont {E.~V.}\
  \bibnamefont {Gryzlova}}, \bibinfo {author} {\bibfnamefont {S.~I.}\
  \bibnamefont {Strakhova}}, \bibinfo {author} {\bibfnamefont {K.}~\bibnamefont
  {Bartschat}}, \bibinfo {author} {\bibfnamefont {N.}~\bibnamefont {Douguet}},
  \bibinfo {author} {\bibfnamefont {J.}~\bibnamefont {Venzke}}, \bibinfo
  {author} {\bibfnamefont {D.}~\bibnamefont {Iablonskyi}}, \bibinfo {author}
  {\bibfnamefont {Y.}~\bibnamefont {Kumagai}}, \bibinfo {author} {\bibfnamefont
  {T.}~\bibnamefont {Takanashi}}, \bibinfo {author} {\bibfnamefont
  {K.}~\bibnamefont {Ueda}}, \bibinfo {author} {\bibfnamefont {A.}~\bibnamefont
  {Fischer}}, \bibinfo {author} {\bibfnamefont {M.}~\bibnamefont {Coreno}},
  \bibinfo {author} {\bibfnamefont {F.}~\bibnamefont {Stienkemeier}}, \bibinfo
  {author} {\bibfnamefont {Y.}~\bibnamefont {Ovcharenko}}, \bibinfo {author}
  {\bibfnamefont {T.}~\bibnamefont {Mazza}}, \ and\ \bibinfo {author}
  {\bibfnamefont {M.}~\bibnamefont {Meyer}},\ }\href {\doibase
  10.1038/nphoton.2016.13} {\bibfield  {journal} {\bibinfo  {journal} {Nat.
  Photonics}\ }\textbf {\bibinfo {volume} {10}},\ \bibinfo {pages} {176}
  (\bibinfo {year} {2016})}\BibitemShut {NoStop}%
\bibitem [{\citenamefont {\v{Z}itnik}\ \emph {et~al.}(2019)\citenamefont
  {\v{Z}itnik}, \citenamefont {Miheli\v{c}}, \citenamefont {Bu\v{c}ar},
  \citenamefont {Hrast}, \citenamefont {Barba}, \citenamefont {Kru\v{s}i\v{c}},
  \citenamefont {Rebernik~Ribi\v{c}}, \citenamefont {Urban\v{c}i\v{c}},
  \citenamefont {Ressel}, \citenamefont {Stupar}, \citenamefont {Poletto},
  \citenamefont {Coreno}, \citenamefont {Gauthier},\ and\ \citenamefont
  {De~Ninno}}]{zitnik:19}%
  \BibitemOpen
  \bibfield  {author} {\bibinfo {author} {\bibfnamefont {M.}~\bibnamefont
  {\v{Z}itnik}}, \bibinfo {author} {\bibfnamefont {A.}~\bibnamefont
  {Miheli\v{c}}}, \bibinfo {author} {\bibfnamefont {K.}~\bibnamefont
  {Bu\v{c}ar}}, \bibinfo {author} {\bibfnamefont {M.}~\bibnamefont {Hrast}},
  \bibinfo {author} {\bibfnamefont {{\v{Z}}.}~\bibnamefont {Barba}}, \bibinfo
  {author} {\bibfnamefont {{\v{S}}.}~\bibnamefont {Kru\v{s}i\v{c}}}, \bibinfo
  {author} {\bibfnamefont {P.}~\bibnamefont {Rebernik~Ribi\v{c}}}, \bibinfo
  {author} {\bibfnamefont {J.}~\bibnamefont {Urban\v{c}i\v{c}}}, \bibinfo
  {author} {\bibfnamefont {B.}~\bibnamefont {Ressel}}, \bibinfo {author}
  {\bibfnamefont {M.}~\bibnamefont {Stupar}}, \bibinfo {author} {\bibfnamefont
  {L.}~\bibnamefont {Poletto}}, \bibinfo {author} {\bibfnamefont
  {M.}~\bibnamefont {Coreno}}, \bibinfo {author} {\bibfnamefont
  {D.}~\bibnamefont {Gauthier}}, \ and\ \bibinfo {author} {\bibfnamefont
  {G.}~\bibnamefont {De~Ninno}},\ }\href {\doibase 10.1103/PhysRevA.99.053423}
  {\bibfield  {journal} {\bibinfo  {journal} {Phys. Rev. A}\ }\textbf {\bibinfo
  {volume} {99}},\ \bibinfo {pages} {053423} (\bibinfo {year}
  {2019})}\BibitemShut {NoStop}%
\bibitem [{\citenamefont {De~Ninno}\ \emph {et~al.}(2020)\citenamefont
  {De~Ninno}, \citenamefont {W\"{a}tzel}, \citenamefont {Ribi\v{c}},
  \citenamefont {Allaria}, \citenamefont {Coreno}, \citenamefont {Danailov},
  \citenamefont {David}, \citenamefont {Demidovich}, \citenamefont {Di~Fraia},
  \citenamefont {Giannessi}, \citenamefont {Hansen}, \citenamefont
  {Kru\v{s}i\v{c}}, \citenamefont {Manfredda}, \citenamefont {Meyer},
  \citenamefont {Miheli\v{c}}, \citenamefont {Mirian}, \citenamefont {Plekan},
  \citenamefont {Ressel}, \citenamefont {R\"{o}sner}, \citenamefont {Simoncig},
  \citenamefont {Spampinati}, \citenamefont {Stupar}, \citenamefont
  {\v{Z}itnik}, \citenamefont {Zangrando}, \citenamefont {Callegari},\ and\
  \citenamefont {Berakdar}}]{deninno:20}%
  \BibitemOpen
  \bibfield  {author} {\bibinfo {author} {\bibfnamefont {G.}~\bibnamefont
  {De~Ninno}}, \bibinfo {author} {\bibfnamefont {J.}~\bibnamefont
  {W\"{a}tzel}}, \bibinfo {author} {\bibfnamefont {P.~R.}\ \bibnamefont
  {Ribi\v{c}}}, \bibinfo {author} {\bibfnamefont {E.}~\bibnamefont {Allaria}},
  \bibinfo {author} {\bibfnamefont {M.}~\bibnamefont {Coreno}}, \bibinfo
  {author} {\bibfnamefont {M.~B.}\ \bibnamefont {Danailov}}, \bibinfo {author}
  {\bibfnamefont {C.}~\bibnamefont {David}}, \bibinfo {author} {\bibfnamefont
  {A.}~\bibnamefont {Demidovich}}, \bibinfo {author} {\bibfnamefont
  {M.}~\bibnamefont {Di~Fraia}}, \bibinfo {author} {\bibfnamefont
  {L.}~\bibnamefont {Giannessi}}, \bibinfo {author} {\bibfnamefont
  {K.}~\bibnamefont {Hansen}}, \bibinfo {author} {\bibfnamefont
  {{\v{S}}.}~\bibnamefont {Kru\v{s}i\v{c}}}, \bibinfo {author} {\bibfnamefont
  {M.}~\bibnamefont {Manfredda}}, \bibinfo {author} {\bibfnamefont
  {M.}~\bibnamefont {Meyer}}, \bibinfo {author} {\bibfnamefont
  {A.}~\bibnamefont {Miheli\v{c}}}, \bibinfo {author} {\bibfnamefont
  {N.}~\bibnamefont {Mirian}}, \bibinfo {author} {\bibfnamefont
  {O.}~\bibnamefont {Plekan}}, \bibinfo {author} {\bibfnamefont
  {B.}~\bibnamefont {Ressel}}, \bibinfo {author} {\bibfnamefont
  {B.}~\bibnamefont {R\"{o}sner}}, \bibinfo {author} {\bibfnamefont
  {A.}~\bibnamefont {Simoncig}}, \bibinfo {author} {\bibfnamefont
  {S.}~\bibnamefont {Spampinati}}, \bibinfo {author} {\bibfnamefont
  {M.}~\bibnamefont {Stupar}}, \bibinfo {author} {\bibfnamefont
  {M.}~\bibnamefont {\v{Z}itnik}}, \bibinfo {author} {\bibfnamefont
  {M.}~\bibnamefont {Zangrando}}, \bibinfo {author} {\bibfnamefont
  {C.}~\bibnamefont {Callegari}}, \ and\ \bibinfo {author} {\bibfnamefont
  {J.}~\bibnamefont {Berakdar}},\ }\href {\doibase 10.1038/s41566-020-0669-y}
  {\bibfield  {journal} {\bibinfo  {journal} {Nat. Photonics}\ }\textbf
  {\bibinfo {volume} {14}},\ \bibinfo {pages} {554} (\bibinfo {year}
  {2020})}\BibitemShut {NoStop}%
\bibitem [{\citenamefont {You}\ \emph {et~al.}(2020)\citenamefont {You},
  \citenamefont {Ueda}, \citenamefont {Gryzlova}, \citenamefont
  {Grum-Grzhimailo}, \citenamefont {Popova}, \citenamefont {Staroselskaya},
  \citenamefont {Tugs}, \citenamefont {Orimo}, \citenamefont {Sato},
  \citenamefont {Ishikawa}, \citenamefont {Carpeggiani}, \citenamefont
  {Csizmadia}, \citenamefont {F\"ule}, \citenamefont {Sansone}, \citenamefont
  {Maroju}, \citenamefont {D'Elia}, \citenamefont {Mazza}, \citenamefont
  {Meyer}, \citenamefont {Callegari}, \citenamefont {Di~Fraia}, \citenamefont
  {Plekan}, \citenamefont {Richter}, \citenamefont {Giannessi}, \citenamefont
  {Allaria}, \citenamefont {De~Ninno}, \citenamefont {Trov\`o}, \citenamefont
  {Badano}, \citenamefont {Diviacco}, \citenamefont {Gaio}, \citenamefont
  {Gauthier}, \citenamefont {Mirian}, \citenamefont {Penco}, \citenamefont
  {Ribi\ifmmode~\check{c}\else \v{c}\fi{}}, \citenamefont {Spampinati},
  \citenamefont {Spezzani},\ and\ \citenamefont {Prince}}]{you:20}%
  \BibitemOpen
  \bibfield  {author} {\bibinfo {author} {\bibfnamefont {D.}~\bibnamefont
  {You}}, \bibinfo {author} {\bibfnamefont {K.}~\bibnamefont {Ueda}}, \bibinfo
  {author} {\bibfnamefont {E.~V.}\ \bibnamefont {Gryzlova}}, \bibinfo {author}
  {\bibfnamefont {A.~N.}\ \bibnamefont {Grum-Grzhimailo}}, \bibinfo {author}
  {\bibfnamefont {M.~M.}\ \bibnamefont {Popova}}, \bibinfo {author}
  {\bibfnamefont {E.~I.}\ \bibnamefont {Staroselskaya}}, \bibinfo {author}
  {\bibfnamefont {O.}~\bibnamefont {Tugs}}, \bibinfo {author} {\bibfnamefont
  {Y.}~\bibnamefont {Orimo}}, \bibinfo {author} {\bibfnamefont
  {T.}~\bibnamefont {Sato}}, \bibinfo {author} {\bibfnamefont {K.~L.}\
  \bibnamefont {Ishikawa}}, \bibinfo {author} {\bibfnamefont {P.~A.}\
  \bibnamefont {Carpeggiani}}, \bibinfo {author} {\bibfnamefont
  {T.}~\bibnamefont {Csizmadia}}, \bibinfo {author} {\bibfnamefont
  {M.}~\bibnamefont {F\"ule}}, \bibinfo {author} {\bibfnamefont
  {G.}~\bibnamefont {Sansone}}, \bibinfo {author} {\bibfnamefont {P.~K.}\
  \bibnamefont {Maroju}}, \bibinfo {author} {\bibfnamefont {A.}~\bibnamefont
  {D'Elia}}, \bibinfo {author} {\bibfnamefont {T.}~\bibnamefont {Mazza}},
  \bibinfo {author} {\bibfnamefont {M.}~\bibnamefont {Meyer}}, \bibinfo
  {author} {\bibfnamefont {C.}~\bibnamefont {Callegari}}, \bibinfo {author}
  {\bibfnamefont {M.}~\bibnamefont {Di~Fraia}}, \bibinfo {author}
  {\bibfnamefont {O.}~\bibnamefont {Plekan}}, \bibinfo {author} {\bibfnamefont
  {R.}~\bibnamefont {Richter}}, \bibinfo {author} {\bibfnamefont
  {L.}~\bibnamefont {Giannessi}}, \bibinfo {author} {\bibfnamefont
  {E.}~\bibnamefont {Allaria}}, \bibinfo {author} {\bibfnamefont
  {G.}~\bibnamefont {De~Ninno}}, \bibinfo {author} {\bibfnamefont
  {M.}~\bibnamefont {Trov\`o}}, \bibinfo {author} {\bibfnamefont
  {L.}~\bibnamefont {Badano}}, \bibinfo {author} {\bibfnamefont
  {B.}~\bibnamefont {Diviacco}}, \bibinfo {author} {\bibfnamefont
  {G.}~\bibnamefont {Gaio}}, \bibinfo {author} {\bibfnamefont {D.}~\bibnamefont
  {Gauthier}}, \bibinfo {author} {\bibfnamefont {N.}~\bibnamefont {Mirian}},
  \bibinfo {author} {\bibfnamefont {G.}~\bibnamefont {Penco}}, \bibinfo
  {author} {\bibfnamefont {P.~c. v.~R.}\ \bibnamefont
  {Ribi\ifmmode~\check{c}\else \v{c}\fi{}}}, \bibinfo {author} {\bibfnamefont
  {S.}~\bibnamefont {Spampinati}}, \bibinfo {author} {\bibfnamefont
  {C.}~\bibnamefont {Spezzani}}, \ and\ \bibinfo {author} {\bibfnamefont
  {K.~C.}\ \bibnamefont {Prince}},\ }\href {\doibase
  10.1103/PhysRevX.10.031070} {\bibfield  {journal} {\bibinfo  {journal} {Phys.
  Rev. X}\ }\textbf {\bibinfo {volume} {10}},\ \bibinfo {pages} {031070}
  (\bibinfo {year} {2020})}\BibitemShut {NoStop}%
\bibitem [{\citenamefont {McCurdy}\ and\ \citenamefont
  {Mart\'{\i}n}(2004)}]{mccurdy:04}%
  \BibitemOpen
  \bibfield  {author} {\bibinfo {author} {\bibfnamefont {C.~W.}\ \bibnamefont
  {McCurdy}}\ and\ \bibinfo {author} {\bibfnamefont {F.}~\bibnamefont
  {Mart\'{\i}n}},\ }\href {\doibase 10.1088/0953-4075/37/4/017} {\bibfield
  {journal} {\bibinfo  {journal} {J. Phys. B}\ }\textbf {\bibinfo {volume}
  {37}},\ \bibinfo {pages} {917} (\bibinfo {year} {2004})}\BibitemShut
  {NoStop}%
\bibitem [{\citenamefont {McCurdy}\ \emph
  {et~al.}(2004{\natexlab{a}})\citenamefont {McCurdy}, \citenamefont
  {Baertschy},\ and\ \citenamefont {Rescigno}}]{mccurdy:04a}%
  \BibitemOpen
  \bibfield  {author} {\bibinfo {author} {\bibfnamefont {C.~W.}\ \bibnamefont
  {McCurdy}}, \bibinfo {author} {\bibfnamefont {M.}~\bibnamefont {Baertschy}},
  \ and\ \bibinfo {author} {\bibfnamefont {T.~N.}\ \bibnamefont {Rescigno}},\
  }\href {\doibase 10.1088/0953-4075/37/17/R01} {\bibfield  {journal} {\bibinfo
   {journal} {J. Phys. B}\ }\textbf {\bibinfo {volume} {37}},\ \bibinfo {pages}
  {R137} (\bibinfo {year} {2004}{\natexlab{a}})}\BibitemShut {NoStop}%
\bibitem [{\citenamefont {McCurdy}\ \emph
  {et~al.}(2004{\natexlab{b}})\citenamefont {McCurdy}, \citenamefont {Horner},
  \citenamefont {Rescigno},\ and\ \citenamefont {Mart\'{\i}n}}]{mccurdy:04b}%
  \BibitemOpen
  \bibfield  {author} {\bibinfo {author} {\bibfnamefont {C.~W.}\ \bibnamefont
  {McCurdy}}, \bibinfo {author} {\bibfnamefont {D.~A.}\ \bibnamefont {Horner}},
  \bibinfo {author} {\bibfnamefont {T.~N.}\ \bibnamefont {Rescigno}}, \ and\
  \bibinfo {author} {\bibfnamefont {F.}~\bibnamefont {Mart\'{\i}n}},\ }\href
  {\doibase 10.1103/PhysRevA.69.032707} {\bibfield  {journal} {\bibinfo
  {journal} {Phys. Rev. A}\ }\textbf {\bibinfo {volume} {69}},\ \bibinfo
  {pages} {032707} (\bibinfo {year} {2004}{\natexlab{b}})}\BibitemShut
  {NoStop}%
\bibitem [{\citenamefont {Horner}\ \emph {et~al.}(2007)\citenamefont {Horner},
  \citenamefont {Morales}, \citenamefont {Rescigno}, \citenamefont
  {Mart\'{\i}n},\ and\ \citenamefont {McCurdy}}]{horner:07}%
  \BibitemOpen
  \bibfield  {author} {\bibinfo {author} {\bibfnamefont {D.~A.}\ \bibnamefont
  {Horner}}, \bibinfo {author} {\bibfnamefont {F.}~\bibnamefont {Morales}},
  \bibinfo {author} {\bibfnamefont {T.~N.}\ \bibnamefont {Rescigno}}, \bibinfo
  {author} {\bibfnamefont {F.}~\bibnamefont {Mart\'{\i}n}}, \ and\ \bibinfo
  {author} {\bibfnamefont {C.~W.}\ \bibnamefont {McCurdy}},\ }\href {\doibase
  10.1103/PhysRevA.76.030701} {\bibfield  {journal} {\bibinfo  {journal} {Phys.
  Rev. A}\ }\textbf {\bibinfo {volume} {76}},\ \bibinfo {pages} {030701(R)}
  (\bibinfo {year} {2007})}\BibitemShut {NoStop}%
\bibitem [{\citenamefont {Miheli\v{c}}(2018)}]{mihelic:18}%
  \BibitemOpen
  \bibfield  {author} {\bibinfo {author} {\bibfnamefont {A.}~\bibnamefont
  {Miheli\v{c}}},\ }\href {\doibase 10.1103/PhysRevA.98.023409} {\bibfield
  {journal} {\bibinfo  {journal} {Phys. Rev. A}\ }\textbf {\bibinfo {volume}
  {98}},\ \bibinfo {pages} {023409} (\bibinfo {year} {2018})}\BibitemShut
  {NoStop}%
\bibitem [{\citenamefont {Scrinzi}(2010)}]{scrinzi:10}%
  \BibitemOpen
  \bibfield  {author} {\bibinfo {author} {\bibfnamefont {A.}~\bibnamefont
  {Scrinzi}},\ }\href {\doibase 10.1103/PhysRevA.81.053845} {\bibfield
  {journal} {\bibinfo  {journal} {Phys. Rev. A}\ }\textbf {\bibinfo {volume}
  {81}},\ \bibinfo {pages} {053845} (\bibinfo {year} {2010})}\BibitemShut
  {NoStop}%
\bibitem [{\citenamefont {Palacios}\ \emph {et~al.}(2007)\citenamefont
  {Palacios}, \citenamefont {McCurdy},\ and\ \citenamefont
  {Rescigno}}]{palacios:07}%
  \BibitemOpen
  \bibfield  {author} {\bibinfo {author} {\bibfnamefont {A.}~\bibnamefont
  {Palacios}}, \bibinfo {author} {\bibfnamefont {C.~W.}\ \bibnamefont
  {McCurdy}}, \ and\ \bibinfo {author} {\bibfnamefont {T.~N.}\ \bibnamefont
  {Rescigno}},\ }\href {\doibase 10.1103/PhysRevA.76.043420} {\bibfield
  {journal} {\bibinfo  {journal} {Phys. Rev. A}\ }\textbf {\bibinfo {volume}
  {76}},\ \bibinfo {pages} {043420} (\bibinfo {year} {2007})}\BibitemShut
  {NoStop}%
\bibitem [{\citenamefont {Palacios}\ \emph {et~al.}(2008)\citenamefont
  {Palacios}, \citenamefont {Rescigno},\ and\ \citenamefont
  {McCurdy}}]{palacios:08}%
  \BibitemOpen
  \bibfield  {author} {\bibinfo {author} {\bibfnamefont {A.}~\bibnamefont
  {Palacios}}, \bibinfo {author} {\bibfnamefont {T.~N.}\ \bibnamefont
  {Rescigno}}, \ and\ \bibinfo {author} {\bibfnamefont {C.~W.}\ \bibnamefont
  {McCurdy}},\ }\href {\doibase 10.1103/PhysRevA.77.032716} {\bibfield
  {journal} {\bibinfo  {journal} {Phys. Rev. A}\ }\textbf {\bibinfo {volume}
  {77}},\ \bibinfo {pages} {032716} (\bibinfo {year} {2008})}\BibitemShut
  {NoStop}%
\bibitem [{\citenamefont {Palacios}\ \emph {et~al.}(2009)\citenamefont
  {Palacios}, \citenamefont {Rescigno},\ and\ \citenamefont
  {McCurdy}}]{palacios:09}%
  \BibitemOpen
  \bibfield  {author} {\bibinfo {author} {\bibfnamefont {A.}~\bibnamefont
  {Palacios}}, \bibinfo {author} {\bibfnamefont {T.~N.}\ \bibnamefont
  {Rescigno}}, \ and\ \bibinfo {author} {\bibfnamefont {C.~W.}\ \bibnamefont
  {McCurdy}},\ }\href {\doibase 10.1103/PhysRevA.79.033402} {\bibfield
  {journal} {\bibinfo  {journal} {Phys. Rev. A}\ }\textbf {\bibinfo {volume}
  {79}},\ \bibinfo {pages} {033402} (\bibinfo {year} {2009})}\BibitemShut
  {NoStop}%
\bibitem [{\citenamefont {Boll}\ \emph {et~al.}(2019)\citenamefont {Boll},
  \citenamefont {Foj\'on}, \citenamefont {McCurdy},\ and\ \citenamefont
  {Palacios}}]{boll:19}%
  \BibitemOpen
  \bibfield  {author} {\bibinfo {author} {\bibfnamefont {D.~I.~R.}\
  \bibnamefont {Boll}}, \bibinfo {author} {\bibfnamefont {O.~A.}\ \bibnamefont
  {Foj\'on}}, \bibinfo {author} {\bibfnamefont {C.~W.}\ \bibnamefont
  {McCurdy}}, \ and\ \bibinfo {author} {\bibfnamefont {A.}~\bibnamefont
  {Palacios}},\ }\href {\doibase 10.1103/PhysRevA.99.023416} {\bibfield
  {journal} {\bibinfo  {journal} {Phys. Rev. A}\ }\textbf {\bibinfo {volume}
  {99}},\ \bibinfo {pages} {023416} (\bibinfo {year} {2019})}\BibitemShut
  {NoStop}%
\bibitem [{\citenamefont {Tao}\ and\ \citenamefont {Scrinzi}(2012)}]{tao:12}%
  \BibitemOpen
  \bibfield  {author} {\bibinfo {author} {\bibfnamefont {L.}~\bibnamefont
  {Tao}}\ and\ \bibinfo {author} {\bibfnamefont {A.}~\bibnamefont {Scrinzi}},\
  }\href {\doibase 10.1088/1367-2630/14/1/013021} {\bibfield  {journal}
  {\bibinfo  {journal} {New J. Phys}\ }\textbf {\bibinfo {volume} {14}},\
  \bibinfo {pages} {013021} (\bibinfo {year} {2012})}\BibitemShut {NoStop}%
\bibitem [{\citenamefont {Scrinzi}(2012)}]{scrinzi:12}%
  \BibitemOpen
  \bibfield  {author} {\bibinfo {author} {\bibfnamefont {A.}~\bibnamefont
  {Scrinzi}},\ }\href {\doibase 10.1088/1367-2630/14/8/085008} {\bibfield
  {journal} {\bibinfo  {journal} {New J. Phys.}\ }\textbf {\bibinfo {volume}
  {14}},\ \bibinfo {pages} {085008} (\bibinfo {year} {2012})}\BibitemShut
  {NoStop}%
\bibitem [{\citenamefont {Sato}\ \emph {et~al.}(2016)\citenamefont {Sato},
  \citenamefont {Ishikawa}, \citenamefont {B\v{r}ezinov\'a}, \citenamefont
  {Lackner}, \citenamefont {Nagele},\ and\ \citenamefont
  {Burgd\"orfer}}]{sato:16}%
  \BibitemOpen
  \bibfield  {author} {\bibinfo {author} {\bibfnamefont {T.}~\bibnamefont
  {Sato}}, \bibinfo {author} {\bibfnamefont {K.~L.}\ \bibnamefont {Ishikawa}},
  \bibinfo {author} {\bibfnamefont {I.}~\bibnamefont {B\v{r}ezinov\'a}},
  \bibinfo {author} {\bibfnamefont {F.}~\bibnamefont {Lackner}}, \bibinfo
  {author} {\bibfnamefont {S.}~\bibnamefont {Nagele}}, \ and\ \bibinfo {author}
  {\bibfnamefont {J.}~\bibnamefont {Burgd\"orfer}},\ }\href {\doibase
  10.1103/PhysRevA.94.023405} {\bibfield  {journal} {\bibinfo  {journal} {Phys.
  Rev. A}\ }\textbf {\bibinfo {volume} {94}},\ \bibinfo {pages} {023405}
  (\bibinfo {year} {2016})}\BibitemShut {NoStop}%
\bibitem [{\citenamefont {Orimo}\ \emph {et~al.}(2018)\citenamefont {Orimo},
  \citenamefont {Sato}, \citenamefont {Scrinzi},\ and\ \citenamefont
  {Ishikawa}}]{orimo:18}%
  \BibitemOpen
  \bibfield  {author} {\bibinfo {author} {\bibfnamefont {Y.}~\bibnamefont
  {Orimo}}, \bibinfo {author} {\bibfnamefont {T.}~\bibnamefont {Sato}},
  \bibinfo {author} {\bibfnamefont {A.}~\bibnamefont {Scrinzi}}, \ and\
  \bibinfo {author} {\bibfnamefont {K.~L.}\ \bibnamefont {Ishikawa}},\ }\href
  {\doibase 10.1103/PhysRevA.97.023423} {\bibfield  {journal} {\bibinfo
  {journal} {Phys. Rev. A}\ }\textbf {\bibinfo {volume} {97}},\ \bibinfo
  {pages} {023423} (\bibinfo {year} {2018})}\BibitemShut {NoStop}%
\bibitem [{\citenamefont {Horner}\ \emph
  {et~al.}(2008{\natexlab{a}})\citenamefont {Horner}, \citenamefont {McCurdy},\
  and\ \citenamefont {Rescigno}}]{horner:08a}%
  \BibitemOpen
  \bibfield  {author} {\bibinfo {author} {\bibfnamefont {D.~A.}\ \bibnamefont
  {Horner}}, \bibinfo {author} {\bibfnamefont {C.~W.}\ \bibnamefont {McCurdy}},
  \ and\ \bibinfo {author} {\bibfnamefont {T.~N.}\ \bibnamefont {Rescigno}},\
  }\href {\doibase 10.1103/PhysRevA.78.043416} {\bibfield  {journal} {\bibinfo
  {journal} {Phys. Rev. A}\ }\textbf {\bibinfo {volume} {78}},\ \bibinfo
  {pages} {043416} (\bibinfo {year} {2008}{\natexlab{a}})}\BibitemShut
  {NoStop}%
\bibitem [{\citenamefont {Horner}\ \emph
  {et~al.}(2008{\natexlab{b}})\citenamefont {Horner}, \citenamefont
  {Rescigno},\ and\ \citenamefont {McCurdy}}]{horner:08b}%
  \BibitemOpen
  \bibfield  {author} {\bibinfo {author} {\bibfnamefont {D.~A.}\ \bibnamefont
  {Horner}}, \bibinfo {author} {\bibfnamefont {T.~N.}\ \bibnamefont
  {Rescigno}}, \ and\ \bibinfo {author} {\bibfnamefont {C.~W.}\ \bibnamefont
  {McCurdy}},\ }\href {\doibase 10.1103/PhysRevA.77.030703} {\bibfield
  {journal} {\bibinfo  {journal} {Phys. Rev. A}\ }\textbf {\bibinfo {volume}
  {77}},\ \bibinfo {pages} {030703(R)} (\bibinfo {year}
  {2008}{\natexlab{b}})}\BibitemShut {NoStop}%
\bibitem [{\citenamefont {Joachain}(1975)}]{joachain:75}%
  \BibitemOpen
  \bibfield  {author} {\bibinfo {author} {\bibfnamefont {C.~J.}\ \bibnamefont
  {Joachain}},\ }\href@noop {} {\emph {\bibinfo {title} {Quantum Collision
  Theory}}}\ (\bibinfo  {publisher} {North-Holland},\ \bibinfo {address}
  {Amsterdam},\ \bibinfo {year} {1975})\BibitemShut {NoStop}%
\bibitem [{\citenamefont {Adawi}(1964)}]{adawi:64}%
  \BibitemOpen
  \bibfield  {author} {\bibinfo {author} {\bibfnamefont {I.}~\bibnamefont
  {Adawi}},\ }\href {\doibase 10.1119/1.1970179} {\bibfield  {journal}
  {\bibinfo  {journal} {Am. J. Phys.}\ }\textbf {\bibinfo {volume} {32}},\
  \bibinfo {pages} {211} (\bibinfo {year} {1964})}\BibitemShut {NoStop}%
\bibitem [{\citenamefont {Olver}\ \emph {et~al.}(2010)\citenamefont {Olver},
  \citenamefont {Lozier}, \citenamefont {F.},\ and\ \citenamefont
  {Clark}}]{olver:10}%
  \BibitemOpen
  \bibinfo {editor} {\bibfnamefont {F.~W.~J.}\ \bibnamefont {Olver}}, \bibinfo
  {editor} {\bibfnamefont {D.~W.}\ \bibnamefont {Lozier}}, \bibinfo {editor}
  {\bibfnamefont {B.~R.}\ \bibnamefont {F.}}, \ and\ \bibinfo {editor}
  {\bibfnamefont {C.~W.}\ \bibnamefont {Clark}},\ eds.,\ \href@noop {} {\emph
  {\bibinfo {title} {NIST Handbook of Mathematical Functions}}}\ (\bibinfo
  {publisher} {Cambridge University Press},\ \bibinfo {address} {New York},\
  \bibinfo {year} {2010})\BibitemShut {NoStop}%
\bibitem [{\citenamefont {Proulx}\ \emph {et~al.}(1994)\citenamefont {Proulx},
  \citenamefont {Pont},\ and\ \citenamefont {Shakeshaft}}]{proulx:94}%
  \BibitemOpen
  \bibfield  {author} {\bibinfo {author} {\bibfnamefont {D.}~\bibnamefont
  {Proulx}}, \bibinfo {author} {\bibfnamefont {M.}~\bibnamefont {Pont}}, \ and\
  \bibinfo {author} {\bibfnamefont {R.}~\bibnamefont {Shakeshaft}},\ }\href
  {\doibase 10.1103/PhysRevA.49.1208} {\bibfield  {journal} {\bibinfo
  {journal} {Phys. Rev. A}\ }\textbf {\bibinfo {volume} {49}},\ \bibinfo
  {pages} {1208} (\bibinfo {year} {1994})}\BibitemShut {NoStop}%
\bibitem [{\citenamefont {Marante}\ \emph {et~al.}(2014)\citenamefont
  {Marante}, \citenamefont {Argenti},\ and\ \citenamefont
  {Mart\'{\i}n}}]{marante:14}%
  \BibitemOpen
  \bibfield  {author} {\bibinfo {author} {\bibfnamefont {C.}~\bibnamefont
  {Marante}}, \bibinfo {author} {\bibfnamefont {L.}~\bibnamefont {Argenti}}, \
  and\ \bibinfo {author} {\bibfnamefont {F.}~\bibnamefont {Mart\'{\i}n}},\
  }\href {\doibase 10.1103/PhysRevA.90.012506} {\bibfield  {journal} {\bibinfo
  {journal} {Phys. Rev. A}\ }\textbf {\bibinfo {volume} {90}},\ \bibinfo
  {pages} {012506} (\bibinfo {year} {2014})}\BibitemShut {NoStop}%
\bibitem [{\citenamefont {Jim\'enez-Gal\'an}\ \emph {et~al.}(2016)\citenamefont
  {Jim\'enez-Gal\'an}, \citenamefont {Mart\'{\i}n},\ and\ \citenamefont
  {Argenti}}]{jimenez:16}%
  \BibitemOpen
  \bibfield  {author} {\bibinfo {author} {\bibfnamefont {A.}~\bibnamefont
  {Jim\'enez-Gal\'an}}, \bibinfo {author} {\bibfnamefont {F.}~\bibnamefont
  {Mart\'{\i}n}}, \ and\ \bibinfo {author} {\bibfnamefont {L.}~\bibnamefont
  {Argenti}},\ }\href {\doibase 10.1103/PhysRevA.93.023429} {\bibfield
  {journal} {\bibinfo  {journal} {Phys. Rev. A}\ }\textbf {\bibinfo {volume}
  {93}},\ \bibinfo {pages} {023429} (\bibinfo {year} {2016})}\BibitemShut
  {NoStop}%
\bibitem [{\citenamefont {Shakeshaft}(2007)}]{shakeshaft:06}%
  \BibitemOpen
  \bibfield  {author} {\bibinfo {author} {\bibfnamefont {R.}~\bibnamefont
  {Shakeshaft}},\ }\href {\doibase 10.1103/PhysRevA.76.063405} {\bibfield
  {journal} {\bibinfo  {journal} {Phys. Rev. A}\ }\textbf {\bibinfo {volume}
  {76}},\ \bibinfo {pages} {063405} (\bibinfo {year} {2007})}\BibitemShut
  {NoStop}%
\bibitem [{\citenamefont {Lambropoulos}\ \emph {et~al.}(1998)\citenamefont
  {Lambropoulos}, \citenamefont {Maragakis},\ and\ \citenamefont
  {Zhang}}]{lambropoulos:98}%
  \BibitemOpen
  \bibfield  {author} {\bibinfo {author} {\bibfnamefont {P.}~\bibnamefont
  {Lambropoulos}}, \bibinfo {author} {\bibfnamefont {P.}~\bibnamefont
  {Maragakis}}, \ and\ \bibinfo {author} {\bibfnamefont {J.}~\bibnamefont
  {Zhang}},\ }\href {\doibase 10.1016/S0370-1573(98)00027-1} {\bibfield
  {journal} {\bibinfo  {journal} {Phys. Rep.}\ }\textbf {\bibinfo {volume}
  {305}},\ \bibinfo {pages} {203 } (\bibinfo {year} {1998})}\BibitemShut
  {NoStop}%
\bibitem [{\citenamefont {Karule}(1988)}]{karule:88}%
  \BibitemOpen
  \bibfield  {author} {\bibinfo {author} {\bibfnamefont {E.}~\bibnamefont
  {Karule}},\ }\href {\doibase 10.1088/0953-4075/21/11/015} {\bibfield
  {journal} {\bibinfo  {journal} {J. Phys. B}\ }\textbf {\bibinfo {volume}
  {21}},\ \bibinfo {pages} {1997} (\bibinfo {year} {1988})}\BibitemShut
  {NoStop}%
\bibitem [{\citenamefont {Karule}(1978)}]{karule:78}%
  \BibitemOpen
  \bibfield  {author} {\bibinfo {author} {\bibfnamefont {E.}~\bibnamefont
  {Karule}},\ }\href {\doibase 10.1088/0022-3700/11/3/015} {\bibfield
  {journal} {\bibinfo  {journal} {J. Phys. B}\ }\textbf {\bibinfo {volume}
  {11}},\ \bibinfo {pages} {441} (\bibinfo {year} {1978})}\BibitemShut
  {NoStop}%
\bibitem [{\citenamefont {Chu}\ and\ \citenamefont {Telnov}(2004)}]{chu:04}%
  \BibitemOpen
  \bibfield  {author} {\bibinfo {author} {\bibfnamefont {S.-I.}\ \bibnamefont
  {Chu}}\ and\ \bibinfo {author} {\bibfnamefont {D.~A.}\ \bibnamefont
  {Telnov}},\ }\href {\doibase 10.1016/j.physrep.2003.10.001} {\bibfield
  {journal} {\bibinfo  {journal} {Phys. Rep.}\ }\textbf {\bibinfo {volume}
  {390}},\ \bibinfo {pages} {1} (\bibinfo {year} {2004})}\BibitemShut {NoStop}%
\bibitem [{\citenamefont {Venuti}\ \emph {et~al.}(1996)\citenamefont {Venuti},
  \citenamefont {Decleva},\ and\ \citenamefont {Lisini}}]{venuti:96}%
  \BibitemOpen
  \bibfield  {author} {\bibinfo {author} {\bibfnamefont {M.}~\bibnamefont
  {Venuti}}, \bibinfo {author} {\bibfnamefont {P.}~\bibnamefont {Decleva}}, \
  and\ \bibinfo {author} {\bibfnamefont {A.}~\bibnamefont {Lisini}},\ }\href
  {\doibase 10.1088/0953-4075/29/22/011} {\bibfield  {journal} {\bibinfo
  {journal} {J. Phys. B}\ }\textbf {\bibinfo {volume} {29}},\ \bibinfo {pages}
  {5315} (\bibinfo {year} {1996})}\BibitemShut {NoStop}%
\bibitem [{\citenamefont {Bachau}\ \emph {et~al.}(2001)\citenamefont {Bachau},
  \citenamefont {Cormier}, \citenamefont {Decleva}, \citenamefont {Hansen},\
  and\ \citenamefont {Mart\'{\i}n}}]{bachau:01}%
  \BibitemOpen
  \bibfield  {author} {\bibinfo {author} {\bibfnamefont {H.}~\bibnamefont
  {Bachau}}, \bibinfo {author} {\bibfnamefont {E.}~\bibnamefont {Cormier}},
  \bibinfo {author} {\bibfnamefont {P.}~\bibnamefont {Decleva}}, \bibinfo
  {author} {\bibfnamefont {J.~E.}\ \bibnamefont {Hansen}}, \ and\ \bibinfo
  {author} {\bibfnamefont {F.}~\bibnamefont {Mart\'{\i}n}},\ }\href {\doibase
  10.1088/0034-4885/64/12/205} {\bibfield  {journal} {\bibinfo  {journal} {Rep.
  Prog. Phys.}\ }\textbf {\bibinfo {volume} {64}},\ \bibinfo {pages} {1815}
  (\bibinfo {year} {2001})}\BibitemShut {NoStop}%
\bibitem [{\citenamefont {Saenz}\ and\ \citenamefont
  {Lambropoulos}(1999)}]{saenz:99}%
  \BibitemOpen
  \bibfield  {author} {\bibinfo {author} {\bibfnamefont {A.}~\bibnamefont
  {Saenz}}\ and\ \bibinfo {author} {\bibfnamefont {P.}~\bibnamefont
  {Lambropoulos}},\ }\href {\doibase 10.1088/0953-4075/32/23/316} {\bibfield
  {journal} {\bibinfo  {journal} {J. Phys. B}\ }\textbf {\bibinfo {volume}
  {32}},\ \bibinfo {pages} {5629} (\bibinfo {year} {1999})}\BibitemShut
  {NoStop}%
\bibitem [{\citenamefont {S\'{a}nchez}\ \emph {et~al.}(1995)\citenamefont
  {S\'{a}nchez}, \citenamefont {Bachau},\ and\ \citenamefont
  {Cormier}}]{sanchez:95}%
  \BibitemOpen
  \bibfield  {author} {\bibinfo {author} {\bibfnamefont {I.}~\bibnamefont
  {S\'{a}nchez}}, \bibinfo {author} {\bibfnamefont {H.}~\bibnamefont {Bachau}},
  \ and\ \bibinfo {author} {\bibfnamefont {E.}~\bibnamefont {Cormier}},\ }\href
  {\doibase 10.1088/0953-4075/28/12/007} {\bibfield  {journal} {\bibinfo
  {journal} {J. Phys. B}\ }\textbf {\bibinfo {volume} {28}},\ \bibinfo {pages}
  {2367} (\bibinfo {year} {1995})}\BibitemShut {NoStop}%
\bibitem [{\citenamefont {Cooper}\ \emph {et~al.}(1963)\citenamefont {Cooper},
  \citenamefont {Fano},\ and\ \citenamefont {Prats}}]{cooper:63}%
  \BibitemOpen
  \bibfield  {author} {\bibinfo {author} {\bibfnamefont {J.~W.}\ \bibnamefont
  {Cooper}}, \bibinfo {author} {\bibfnamefont {U.}~\bibnamefont {Fano}}, \ and\
  \bibinfo {author} {\bibfnamefont {F.}~\bibnamefont {Prats}},\ }\href
  {\doibase 10.1103/PhysRevLett.10.518} {\bibfield  {journal} {\bibinfo
  {journal} {Phys. Rev. Lett.}\ }\textbf {\bibinfo {volume} {10}},\ \bibinfo
  {pages} {518} (\bibinfo {year} {1963})}\BibitemShut {NoStop}%
\bibitem [{\citenamefont {O'Keeffe}\ \emph {et~al.}(2010)\citenamefont
  {O'Keeffe}, \citenamefont {Bolognesi}, \citenamefont {Miheli\v{c}},
  \citenamefont {Moise}, \citenamefont {Richter}, \citenamefont {Cautero},
  \citenamefont {Stebel}, \citenamefont {Sergo}, \citenamefont {Pravica},
  \citenamefont {Ovcharenko}, \citenamefont {Decleva},\ and\ \citenamefont
  {Avaldi}}]{okeeffe:10}%
  \BibitemOpen
  \bibfield  {author} {\bibinfo {author} {\bibfnamefont {P.}~\bibnamefont
  {O'Keeffe}}, \bibinfo {author} {\bibfnamefont {P.}~\bibnamefont {Bolognesi}},
  \bibinfo {author} {\bibfnamefont {A.}~\bibnamefont {Miheli\v{c}}}, \bibinfo
  {author} {\bibfnamefont {A.}~\bibnamefont {Moise}}, \bibinfo {author}
  {\bibfnamefont {R.}~\bibnamefont {Richter}}, \bibinfo {author} {\bibfnamefont
  {G.}~\bibnamefont {Cautero}}, \bibinfo {author} {\bibfnamefont
  {L.}~\bibnamefont {Stebel}}, \bibinfo {author} {\bibfnamefont
  {R.}~\bibnamefont {Sergo}}, \bibinfo {author} {\bibfnamefont
  {L.}~\bibnamefont {Pravica}}, \bibinfo {author} {\bibfnamefont
  {E.}~\bibnamefont {Ovcharenko}}, \bibinfo {author} {\bibfnamefont
  {P.}~\bibnamefont {Decleva}}, \ and\ \bibinfo {author} {\bibfnamefont
  {L.}~\bibnamefont {Avaldi}},\ }\href {\doibase 10.1103/PhysRevA.82.052522}
  {\bibfield  {journal} {\bibinfo  {journal} {Phys. Rev. A}\ }\textbf {\bibinfo
  {volume} {82}},\ \bibinfo {pages} {052522} (\bibinfo {year}
  {2010})}\BibitemShut {NoStop}%
\bibitem [{\citenamefont {O'Keeffe}\ \emph {et~al.}(2013)\citenamefont
  {O'Keeffe}, \citenamefont {Miheli\v{c}}, \citenamefont {Bolognesi},
  \citenamefont {\v{Z}itnik}, \citenamefont {Moise}, \citenamefont {Richter},\
  and\ \citenamefont {Avaldi}}]{okeeffe:13}%
  \BibitemOpen
  \bibfield  {author} {\bibinfo {author} {\bibfnamefont {P.}~\bibnamefont
  {O'Keeffe}}, \bibinfo {author} {\bibfnamefont {A.}~\bibnamefont
  {Miheli\v{c}}}, \bibinfo {author} {\bibfnamefont {P.}~\bibnamefont
  {Bolognesi}}, \bibinfo {author} {\bibfnamefont {M.}~\bibnamefont
  {\v{Z}itnik}}, \bibinfo {author} {\bibfnamefont {A.}~\bibnamefont {Moise}},
  \bibinfo {author} {\bibfnamefont {R.}~\bibnamefont {Richter}}, \ and\
  \bibinfo {author} {\bibfnamefont {L.}~\bibnamefont {Avaldi}},\ }\href
  {\doibase 10.1088/1367-2630/15/1/013023} {\bibfield  {journal} {\bibinfo
  {journal} {New J. Phys.}\ }\textbf {\bibinfo {volume} {15}},\ \bibinfo
  {pages} {013023} (\bibinfo {year} {2013})}\BibitemShut {NoStop}%
\bibitem [{\citenamefont {Carette}\ \emph {et~al.}(2013)\citenamefont
  {Carette}, \citenamefont {Dahlstr\"om}, \citenamefont {Argenti},\ and\
  \citenamefont {Lindroth}}]{carette:13}%
  \BibitemOpen
  \bibfield  {author} {\bibinfo {author} {\bibfnamefont {T.}~\bibnamefont
  {Carette}}, \bibinfo {author} {\bibfnamefont {J.~M.}\ \bibnamefont
  {Dahlstr\"om}}, \bibinfo {author} {\bibfnamefont {L.}~\bibnamefont
  {Argenti}}, \ and\ \bibinfo {author} {\bibfnamefont {E.}~\bibnamefont
  {Lindroth}},\ }\href {\doibase 10.1103/PhysRevA.87.023420} {\bibfield
  {journal} {\bibinfo  {journal} {Phys. Rev. A}\ }\textbf {\bibinfo {volume}
  {87}},\ \bibinfo {pages} {023420} (\bibinfo {year} {2013})}\BibitemShut
  {NoStop}%
\bibitem [{\citenamefont {Brink}\ and\ \citenamefont
  {Satchler}(1975)}]{brink:75}%
  \BibitemOpen
  \bibfield  {author} {\bibinfo {author} {\bibfnamefont {D.~M.}\ \bibnamefont
  {Brink}}\ and\ \bibinfo {author} {\bibfnamefont {G.~R.}\ \bibnamefont
  {Satchler}},\ }\href@noop {} {\emph {\bibinfo {title} {Angular momentum}}},\
  \bibinfo {edition} {2nd}\ ed.\ (\bibinfo  {publisher} {Clarendon},\ \bibinfo
  {address} {Oxford},\ \bibinfo {year} {1975})\BibitemShut {NoStop}%
\end{thebibliography}%

\end{document}